\pdfoutput=1
\documentclass[aps,pre,reprint,floatfix,amsmath,amssymb,showkeys,showpacs]{revtex4-1}
\usepackage[USenglish]{babel}
\usepackage{ucs}				%% unicode support part 1
\usepackage[utf8x]{inputenc}		%% utf8x unicode support part 2
\usepackage{graphicx}			%% include figure files
\usepackage{ifpdf}
\ifpdf
\DeclareGraphicsExtensions{%
    .pdf,.PDF,%
    .png,.PNG,%
    .jpg,.mps,.jpeg,.jbig2,.jb2,.JPG,.JPEG,.JBIG2,.JB2}
\fi

\usepackage[breaklinks=true]{hyperref}			%% clickable links; wrap links e.g. in references
\usepackage{amsmath}
\usepackage{dsfont} %font for real and natural numbers
\usepackage{color}
\usepackage[nointegrals]{wasysym}		%% additional symbols
\usepackage{amssymb}

\usepackage[]{datetime}\newdateformat{mydate}{\THEYEAR-\twodigit{\THEMONTH}-\twodigit{\THEDAY}}
\DeclareMathOperator*{\diag}{diag}
\DeclareMathOperator*{\tr}{tr}
\DeclareMathOperator*{\erf}{erf}
\DeclareMathOperator*{\erfi}{erfi}

\definecolor{darkred}{RGB}{139,0,0}
\definecolor{darkgreen}{RGB}{0,100,0}
\definecolor{darkblue}{RGB}{0,0,139}
\definecolor{darkviolet}{RGB}{148,0,211}

\begin{document}

%%% structural definitions
\newcommand*{\Secref}[1]{Sec.~\ref{#1}}
\newcommand*{\Eqref}[1]{Eq.~(\ref{#1})}
\newcommand*{\EqsToref}[2]{Eqs.~(\ref{#1}) to (\ref{#2})}
\newcommand*{\EqsAndref}[2]{Eqs.~(\ref{#1}) and (\ref{#2})}
\newcommand*{\EqsCommaAndref}[3]{Eqs.~(\ref{#1}), (\ref{#2}) and (\ref{#3})}
\newcommand*{\EqsOrref}[2]{Eq.~(\ref{#1}) or (\ref{#2})}
\newcommand*{\EqsCommaOrref}[3]{Eq.~(\ref{#1}), (\ref{#2}) or (\ref{#3})}
\newcommand*{\Figref}[1]{Fig.~\ref{#1}}
\newcommand*{\Appref}[1]{\appendixname~\ref{#1}}
\newcommand*{\todo}[1]{\PackageWarning{myToDo}{#1}\emph{\color{red}#1}}

%%% generic math definitions
\newcommand*{\intd}[1]{\ensuremath{\text{d}#1}}
\newcommand*{\intdend}[1]{\ensuremath{\;\intd{#1}}}
\newcommand*{\intdbegin}[1]{\ensuremath{\!\intd{#1}}\;}
\newcommand*{\transpose}{\ensuremath{^\mathsf{T}}}
\newcommand*{\imag}{\ensuremath{\text{i}}}
\newcommand*{\avg}[1]{\ensuremath{\left\langle#1\right\rangle}}
\newcommand*{\abs}[1]{\ensuremath{\left\lvert#1\right\rvert}}
\newcommand*{\vektor}[1]{\ensuremath{\boldsymbol{\mathbf{#1}}}}
\newcommand*{\UnitVektor}{\ensuremath{\hat{\vektor{e}}}}
\newcommand*{\mat}[1]{\ensuremath{\boldsymbol{\mathbf{#1}}}}
\newcommand*{\skalar}{\!\cdot\!}
\newcommand*{\dd}{\intd{}}
\newcommand*{\onehalf}{\frac{1}{2}}
\newcommand*{\partderiv}[2]{\frac{\partial #1}{\partial #2}}
\newcommand*{\deriv}[2]{\frac{\intd{#1}}{\intd{#2}}}
\newcommand*{\ten}[1]{{\bf #1}}
\newcommand*{\FourierTransform}[1]{\ensuremath{\mathcal{F}\left[#1\right]}}
\newcommand*{\InverseFourierTransform}[1]{\ensuremath{\mathcal{F}^{-1}\left[#1\right]}}
\newcommand*{\modifiedBessel}[1]{\ensuremath{I_{#1}}}

%%% symbols with pstricks
% \newlength{\myL}
% \newcommand*{\myopencircleParam}[3]{\setlength{\myL}{#1}\psset{unit=1.0}\begin{pspicture}(-1.1\myL,-0.9\myL)(1.1\myL,\myL)\pscircle[linewidth=#2,linecolor=#3](0,0){\myL}\end{pspicture}}
% \newcommand*{\myopenpentagonParam}[3]{\setlength{\myL}{#1}\psset{unit=1.0}\begin{pspicture}(-1.1\myL,-0.9\myL)(1.1\myL,\myL)\psline[linewidth=#2,linecolor=#3](0.5877\myL,-0.8090\myL)(-0.5877\myL,-0.8090\myL)(-0.9511\myL,0.3090\myL)(0,\myL)(0.9511\myL,0.3090\myL)(0.5877\myL,-0.8090\myL)\end{pspicture}}
% \newcommand*{\myopendiamondParam}[3]{\setlength{\myL}{#1}\psset{unit=1.0}\begin{pspicture}(-1.1\myL,-0.9\myL)(1.1\myL,\myL)\psdiamond[linewidth=#2,linecolor=#3](0,0)(\myL,\myL)\end{pspicture}}
% \newcommand*{\myfilledsquareParam}[3]{\setlength{\myL}{#1}\psset{unit=1.0}\begin{pspicture}(0,0)(\myL,\myL)\psframe*[linewidth=#2,linecolor=#3](0,0)(\myL,\myL)\end{pspicture}}
% \newcommand*{\myfilledtriangleParam}[3]{\setlength{\myL}{#1}\psset{unit=1.0}\begin{pspicture}(-\myL,0)(\myL,1.73\myL)\pstriangle*[linewidth=#2,linecolor=#3](0,0)(2.0\myL,1.73\myL)\end{pspicture}}

% \newcommand*{\myopencircle}[1]{{\protect\myopencircleParam{3pt}{0.8pt}{#1}}}
% \newcommand*{\myopendiamond}[1]{{\protect\myopendiamondParam{3pt}{0.8pt}{#1}}}
% \newcommand*{\myopenpentagon}[1]{{\protect\myopenpentagonParam{3pt}{0.8pt}{#1}}}
% \newcommand*{\myfilledsquare}[1]{{\protect\myfilledsquareParam{5pt}{0.8pt}{#1}}}
% \newcommand*{\myfilledtriangle}[1]{{\protect\myfilledtriangleParam{3pt}{0.8pt}{#1}}}

% %%% symbols from package wasysym
\newcommand*{\myopencircle}[1]{{\color{#1}$\ensuremath{\boldsymbol{\Circle}}$}}
\newcommand*{\myopendiamond}[1]{{\color{#1}$\ensuremath{\boldsymbol{\Diamond}}$}}
\newcommand*{\myopenpentagon}[1]{{\color{#1}$\ensuremath{\boldsymbol{\pentagon}}$}}
\newcommand*{\myfilledsquare}[1]{{\color{#1}$\ensuremath{\scriptstyle\blacksquare}$}}
\newcommand*{\myfilledtriangle}[1]{{\color{#1}$\ensuremath{\boldsymbol{\blacktriangle}}$}}

%%% specific math definitions
\newcommand*{\dc}{\ensuremath{D}}	%% diffusion coefficient
\newcommand*{\dcx}{\ensuremath{\dc_{x}}}
\newcommand*{\dcy}{\ensuremath{\dc_{y}}}
\newcommand*{\dcEff}{\ensuremath{\dc'}}
\newcommand*{\meanD}{\ensuremath{\avg{D}}}
\newcommand*{\SingleParticleDiffusivity}{\ensuremath{D_{t}(\tau)}}
\newcommand*{\posv}{\ensuremath{\vektor{x}}}	%% position vector, e.g. trajectory
\newcommand*{\disv}{\ensuremath{\vektor{r}}}  %% displacement vector, e.g. between two position of trajectory
\newcommand*{\dTimesTau}{\ensuremath{d\,\tau}}
\newcommand*{\deltadiffusivitiesdefinition}{\ensuremath{\delta\left[D-\SingleParticleDiffusivity\right]}}
\newcommand*{\tensorDiagD}{\ensuremath{\hat{\ten{D}}}}
\newcommand*{\pdim}[1]{\ensuremath{p^{#1\text{d}}}}
\newcommand*{\piso}[2]{\ensuremath{\pdim{#1}_{#2}}}	%%
\newcommand*{\paniso}[1]{\ensuremath{\pdim{#1}_{\tensorDiagD}}}
\newcommand*{\puni}[1]{\ensuremath{\pdim{#1}_{\text{uni}}}}
\newcommand*{\Giso}[2]{\ensuremath{G^{#1\text{d}}_{#2}}}
\newcommand*{\Ganiso}[1]{\ensuremath{G^{#1\text{d}}_{\tensorDiagD}}}
\newcommand*{\measured}[1]{\ensuremath{\tilde{#1}}}
\newcommand*{\anisotropyMeasure}{\ensuremath{\eta}}
\newcommand*{\submatrixOfSigma}{\ensuremath{\mat{\Sigma}^M_{\vektor{\alpha}}}}	%% war mal: \tilde{\mat{\Sigma}}^M_{\vektor{\alpha}} \mat{\Sigma}^M_\vektor{\alpha}
\newcommand*{\momentm}{\ensuremath{\moment{m}}}
\newcommand*{\moment}[1]{\ensuremath{M_{#1}}}
\newcommand*{\multipleEigenvalueOfD}[1]{\ensuremath{D^{(#1)}}}
\newcommand*{\asymptoticD}{\ensuremath{D_{\infty}}}
\newcommand*{\ParamDegeneracy}{M}
\newcommand*{\vecalphaTwoElements}[2]{\ensuremath{\left(\begin{smallmatrix} #1 & #2 \end{smallmatrix}\right)}}
\newcommand*{\vecalphaOneElement}[1]{\ensuremath{\left(\begin{smallmatrix} #1 \end{smallmatrix}\right)}}
\newcommand*{\effectiveDimensionality}{\ensuremath{N_\text{eff}}}

% \newcommand*{\pos}{\ensuremath{x}}	%% scalar position
% \newcommand*{\dis}{\ensuremath{r}}	%% scalar displacement
% \newcommand*{\disdirection}[1]{\ensuremath{r_{#1}}}	%% displacement in given direction
% \newcommand*{\kvdirection}{\ensuremath{\hat{\vektor{k}}}}	%% k vector for arbitrary direction
% %\newcommand*{\kdirection}[1]{\ensuremath{k_{#1}}}	%% k in given direction
% \newcommand*{\kvpicked}{\ensuremath{(k,0,\dots,0)\transpose}}	%% k vector in direction 1
% \newcommand*{\kvarbitrary}{\ensuremath{k\,\kUnitVektor}}	%% k vector in arbitrary direction
% \newcommand*{\kvabsvalue}{\ensuremath{\lvert\vektor{k}\rvert}}	%% absolute value of vector k
% \newcommand*{\pprojection}{\ensuremath{p_{1}}}	%% projected density
% \newcommand*{\pradial}{\ensuremath{p_{\text{r}}}}	%% radial density
% \newcommand*{\PsiVektor}{\ensuremath{\Psi}}
% \newcommand*{\Psiprojection}{\ensuremath{\Psi_{1}}}	%% projected signal attenuation
% \newcommand*{\Psiradial}{\ensuremath{\tilde{\Psi}}}	%% radial signal attenuation
% \newcommand*{\PsiIsotropic}{\ensuremath{\Psi_{1}}}	%% isotropic signal attenuation
% \newcommand*{\kTilde}{\ensuremath{\tilde{k}}}
% \newcommand*{\pTilde}{\ensuremath{\tilde{p}(D,\kTilde(\tau))}}
% \newcommand*{\kmax}{\ensuremath{k_{\text{max}}}}
% \newcommand*{\GselfVektor}{\ensuremath{G_{\text{s}}(\disv,\tau)}}
% \newcommand*{\GselfSkalar}{\ensuremath{G_{\text{s}}(\dis,\tau)}}
% \newcommand*{\Dik}[1]{\ensuremath{D_{#1} \vektor{k}^2}}
% \newcommand*{\bigO}[1]{\ensuremath{O\left(#1\right)}}%%%\mathcal{O}
% \newcommand*{\iisf}{\ensuremath{S(\vektor{k},\tau)}}

%\preprint{APS/123-QED}

\title{Characterizing \texorpdfstring{$N$}{N}-dimensional anisotropic Brownian motion by the distribution of diffusivities}

\newcommand*{\affiliationTUCxKSND}{Technische Universität Chemnitz, Faculty of Sciences, Institute of Physics, Complex Systems and Nonlinear Dynamics, D-09107 Chemnitz, Germany}
\author{Mario Heidernätsch}
\affiliation{\affiliationTUCxKSND}
\author{Michael Bauer}
\affiliation{\affiliationTUCxKSND}
\author{Günter Radons}%
 \email{radons@physik.tu-chemnitz.de}
\affiliation{\affiliationTUCxKSND}

\date{\today}% It is always \today, today, but any date may be explicitly specified

\begin{abstract}
Anisotropic diffusion processes emerge in various fields such as transport in biological tissue and diffusion in liquid crystals. In such systems, the motion is described by a diffusion tensor. For a proper characterization of processes with more than one diffusion coefficient an average description by the mean squared displacement is often not sufficient. Hence, in this paper, we use the distribution of diffusivities to study diffusion in a homogeneous anisotropic environment. We derive analytical expressions of the distribution and relate its properties to an anisotropy measure based on the mean diffusivity and the asymptotic decay of the distribution. Both quantities are easy to determine from experimental data and reveal the existence of more than one diffusion coefficient, which allows the distinction between isotropic and anisotropic processes. We further discuss the influence on the analysis of projected trajectories, which are typically accessible in experiments. For the experimentally relevant cases of two- and three-dimensional anisotropic diffusion we derive specific expressions, determine the diffusion tensor, characterize the anisotropy, and demonstrate the applicability for simulated trajectories.
\end{abstract}


\maketitle

\section{\label{sec:Introduction}Introduction}

The random motion of suspended particles in a fluid, which is usually referred to as Brownian motion, is an old but still fascinating phenomenon. Especially, when inhomogeneous \cite{vanKampen1988673, Christensen20035171, Bringuier2011} or anisotropic media \cite{schulz2010, schulz2011, PhysRevLett.79.4922} are involved, many questions are still open. From the theoretical point of view, much work has been done \cite{wax1954} to predict the statistical properties of the trajectories of such particles using stochastic methods. On the other side, the development of experiments only recently allows obtaining the paths of individual molecules and particles. Especially the observation of two-dimensional trajectories using video-microscopic methods, for instance by single-particle tracking (SPT), is already successfully applied in biological systems \cite{SaxtonAnnuRef1997, Dahan17102003} or to understand the microrheological properties of complex liquids \cite{PhysRevLett.90.108301, PhysRevLett.79.3282}. But also the observation of three-dimensional paths becomes feasible \cite{NanoLett.7.11, An.Chem.80.24, spille2012}. The statistical analysis of these trajectories is usually accomplished by measuring the mean square displacement (msd) in order to get the diffusion coefficients for the matching theoretical description. However, in the anisotropic case the diffusive properties depend on the direction of motion and are described by a diffusion tensor. In such systems, the analysis of msds turned out to be not sufficient to determine the anisotropy and extract the values of the diffusion coefficients \cite{PhysRevE.75.021112, kaerger2012, hanasaki2012}. For similar reasons, we already introduced the distribution of single-particle diffusivities as an advanced method to analyze stochastic motion in heterogeneous systems \cite{bauer2011} involving more than one diffusion coefficient. It should be noted that this distribution is closely related to the displacement distribution \cite{hellriegel2005, hoefling2013}. However, the distribution of diffusivities is superior since it is stationary for time-homogeneous diffusion processes. Thus, experiments conducted on different time scales can be compared easily. Furthermore, this new method was extended to the distribution of generalized diffusivities to characterize data from anomalous diffusion processes, which offers, for instance, a deeper understanding of weak ergodicity breaking \cite{albers2013}.

In the current article, we show the applicability of the distribution of diffusivities to analyze trajectories of homogeneous anisotropic Brownian motion. We present the properties of the distribution as well as their relations to established quantities. In order to assess the parameters of the process, we calculate the characteristic function, cumulants and moments of the distribution. For the asymptotic decay of the distribution of diffusivities, we derive a general expression, which involves the largest diffusion coefficient of the system. In conjunction with the mean diffusion coefficient of the system, the asymptotic decay enables a data-based distinction between isotropic and anisotropic processes. Based on these quantities, we provide a measure to characterize the anisotropy of the process from the analysis of SPT data. Since in experiments the reconstruction of the complete diffusion tensor is of great interest, we extend our concept to tensorial diffusivities, which offer a simple method to determine the entries of the tensor.

Due to restrictions in SPT experiments the complete trajectory is often not accessible \cite{hellriegel2005, schulz2010}. Hence, we investigate the influence on the distribution of diffusivities and the detection of the anisotropy if only projections of the actual trajectory are observed. Even in such cases it is possible to estimate bounds of the diffusion coefficients from the given projections of the diffusion tensor. Since especially two-dimensional and three-dimensional diffusion processes have a high relevance in experiments we apply our considerations to these systems. For homogeneous anisotropic diffusion in two dimensions an analytical expression of the distribution of diffusivities exists and its moments can be related to the diffusion coefficients, which enter the anisotropy measure. Moreover, we explain the details of reconstructing the diffusion tensor from the tensorial diffusivities as well as from projections of the trajectory. Three-dimensional processes are investigated analogously although a closed-form expression of the distribution of diffusivities does not exist. Additionally, we deal with anisotropic processes where one diffusion coefficient is degenerated corresponding to diffusion of uniaxial molecules typical for liquid crystalline systems \cite{deGennes_B95}.

The paper is organized as follows. In \Secref{sec:Definition}, we briefly recall the theoretical principles of anisotropic Brownian motion based on the diffusion tensor and introduce the distribution of single-particle diffusivities, its properties and relations to established quantities. To apply our new concepts to $N$-dimensional homogeneous anisotropic diffusion processes, we provide in \Secref{sec:properties} a general expression for the distribution of diffusivities. We demonstrate how to distinguish between isotropic and anisotropic processes and explain the reconstruction of the diffusion tensor. Since in experiments typically a projection of the motion is observed we characterize the distribution of diffusivities of the projected trajectories. Finally, in \Secref{sec:Specific systems}, we apply our results to specific systems of anisotropic diffusion which are typical for experimental setups. We substantiate the applicability of our findings by analyzing data from simulated anisotropic diffusion processes.

\section{\label{sec:Definition}Definitions}

\subsection{\label{sec:definitions anisotropic diffusion}Anisotropic diffusion}
An $N$-dimensional anisotropic Brownian motion is completely defined by its propagator \cite{Risken1989}
\begin{align}
    \label{eqn:n-propagator}
    p(\posv, t| \posv' , t') = &  \frac{(2 \pi)^{-\frac{N}{2}}}{\sqrt{[2 (t-t')]^N \det{\ten{D}}}} \nonumber\\
    & \times\exp\left[-\onehalf \frac{1}{2(t-t')}(\posv - \posv')\transpose \ten{D}^{-1} (\posv - \posv')\right]
\end{align}
where $\ten{D} = \ten{O}\transpose\tensorDiagD\ten{O}$ is the positive definite and symmetric diffusion tensor, $\tensorDiagD = \diag(D_1, D_2, \dotsc, D_N)$ denotes its diagonalized form with the diffusion coefficients $D_{i}$ belonging to the principal axes, and $\ten{O}$ is an orthogonal matrix which describes the orientation of the principal axes relative to the frame of reference.

For the simulation of such processes an alternative description exists, where the trajectories are evolved by the Langevin equation
\begin{equation}
    \label{eqn:langevin-nd}
    \deriv{\posv}{t} = \sqrt{ 2 \ten{D} } \vektor{\xi}(t)
\end{equation}
with $\sqrt{\ten{D}} = \ten{O}\transpose\sqrt{\tensorDiagD}\ten{O}$ and $\sqrt{\tensorDiagD} = \diag(\sqrt{D_1}, \sqrt{D_2}, \dotsc, \sqrt{D_N})$. The vector $\vektor{\xi}(t) = [\xi_{1}(t),\dotsc,\xi_{N}(t)]\transpose$ denotes Gaussian white noise in $N$ dimensions with $\avg{\vektor{\xi}(t)} = \vektor{0}$ and $\avg{\xi_i(t) \xi_j(t')} = \delta_{i j}\delta(t-t') \,\forall\, i,j \in \{1,2,\dotsc,N\}$.

%In an alternative description the time evolution of the probability density of $\posv$ is defined by an $N$-dimensional diffusion equation
%\begin{equation}
%\partderiv{p(\posv(t))}{t} = \sum_{i,j = 1}^N D_{ij} \frac{\partial^2 p(\posv(t))}{\partial x_i \partial x_j}
%\end{equation}
%where $x_i$ and $D_{ij}$ are the components of $\posv$ and $\ten{D}$, respectively. \todo{Wozu wird die Fokker-Planck-Gleichung benötigt? Absatz kann eigentlich weggelassen werden. Langevin kann bleiben, da für Simulation verwendet.}

Assuming time-translation invariance \Eqref{eqn:n-propagator} is simplified to the probability density $p(\posv' + \disv, \tau \vert \posv')$ of displacements $\disv = \posv-\posv'$ by substituting $\tau = t-t'$. This conditional probability density is averaged by the equilibrium distribution $p_{0}(\posv')$ given by the Boltzmann distribution to obtain the ensemble-averaged probability density
\begin{align}
\label{eqn:distribution of displacements}
p(\disv,\tau) &= \int \intdbegin{^{N}\posv'} p(\posv' + \disv, \tau \vert \posv') p_{0}(\posv') \nonumber\\
              &= \frac{(2 \pi)^{-\frac{N}{2}}}{\sqrt{\det \mat{\Sigma}}} \exp\left(-\onehalf \disv\transpose\mat{\Sigma}^{-1}\disv\right)
\end{align}
of a displacement $\disv = (r_{1},\dotsc,r_{N})\transpose$ in the time interval $\tau$. Thus, $p(\disv,\tau)$ is an $N$-dimensional Gaussian distribution with zero mean and covariance tensor $\mat{\Sigma} = 2 \tau \ten{D}$.

Expressions with dimensionality $N > 3$ may be interesting for simultaneous diffusion of $d$ particles corresponding to an extended many-particle state space $\posv(\posv_{1}, \dotsc, \posv_{d})$.

\subsection{\label{sec:definitions distribution of diffusivities}Distribution of diffusivities}
By observing a trajectory $\posv(t)$ of an arbitrary stochastic process in $N$ dimensions individual displacements during a given time lag $\tau$ can simply be measured for a certain particle. Moreover, it is natural to relate each displacement to a single-particle diffusivity
\begin{equation}
 \label{eqn:single-particle diffusivity}
 \SingleParticleDiffusivity = \frac{[\posv(t+\tau)-\posv(t)]^{2}}{2N \tau}
\text{.}
\end{equation}
This simple transformation of displacements to diffusivities offers the advantage to compare these quantities for different experimental setups and different $\tau$. Since for a fixed time lag $\tau$ the single-particle diffusivity is fluctuating along a trajectory an important quantity is given by the probability density $p(D)$. Therefore, the distribution of single-particle diffusivities \cite{bauer2011} is defined as
\begin{equation}
\label{eqn:definition of distribution of diffusivities}
p(D,\tau) = \avg{\deltadiffusivitiesdefinition}
\text{,}
\end{equation}
where $\avg{\dotso}$ either denotes a time average $\avg{\dotso} = \lim_{T \to \infty} 1/T \int_{0}^{T} \dotso \intd{t}$, which is typically accessible by SPT, or an ensemble average as measured by other experimental methods, such as nuclear magnetic resonance \cite{price2009}. For ergodic systems, as considered here, time average and ensemble average coincide. It should be noted that other definitions of diffusivity distributions exist in the literature \cite{saxton1997}.

For time-homogeneous systems, i.e., when the distribution of displacements $p(\disv,\tau)$ is independent of $t$, \Eqref{eqn:definition of distribution of diffusivities} can be rewritten as
\begin{equation}
\label{eqn:definition of distribution of diffusivities in stationary case}
p(D,\tau) = \int \intdbegin{^{N}\disv} \delta\left(D - \frac{\disv^2}{2 N \tau}\right) p(\disv,\tau)
\end{equation}
transforming $p(\disv,\tau)$ into the distribution of diffusivities.

For data from SPT experiments, displacements from a trajectory are transformed to diffusivities according to \Eqref{eqn:single-particle diffusivity} and the distribution of diffusivities is obtained by binning these diffusivities into a normalized histogram according to \Eqref{eqn:definition of distribution of diffusivities}.

For homogeneous isotropic processes in $N$ dimensions the msd grows linearly with $\tau$, since it obeys the well-known Einstein relation $\avg{r^2(\tau)} = 2 N D_c \tau$, where $D_c$ is the diffusion coefficient of the process. Due to the transformation of displacements to diffusivities by \Eqref{eqn:single-particle diffusivity} the linear dependence on $\tau$ is removed. Hence, the corresponding distribution of diffusivities becomes stationary and comprises single-particle diffusivities fluctuating around $D_c$. For $N$-dimensional homogeneous isotropic processes the distribution of diffusivities
\begin{equation}
    \label{eqn:dod homogeneous isotropic system in n dimensions}
    \piso{N}{D_c}(D) = \left(\frac{N}{2 D_c}\right)^{\frac{N}{2}} \frac{D^{\frac{N}{2}-1}}{\Gamma(\frac{N}{2})} \exp\left(-\frac{N}{2 D_c} D\right)
\end{equation}
is obtained, where $\Gamma(x)$ denotes the gamma function. This distribution is identified as a $\chi^2$-distribution of $N$ degrees of freedom and results directly from the sum of the squares of $N$ independent and identically distributed Gaussian random variables with variance $D_c/N$ and vanishing mean. Since these variables are the squared and rescaled components of the displacement vector $r_i^2(\tau)/(2 N \tau)$, their sum corresponds to the diffusivity.

For inhomogeneous isotropic diffusion processes which are ergodic \Eqref{eqn:dod homogeneous isotropic system in n dimensions} provides a further useful application. Since for normal diffusion in $N$ dimensions the Einstein relation holds for large $\tau$, $p(D,\tau)$ converges to the stationary distribution given by \Eqref{eqn:dod homogeneous isotropic system in n dimensions}. In this case, $D_c$ is the mean diffusion coefficient of the process.

\subsection{\label{sec:definitions general moments}Moments}
The distribution of diffusivities is fully characterized by its corresponding moments
\begin{equation}
 \label{eqn:moments of dod}
 \momentm(\tau) = \avg{D(\tau)^{m}} = \int\limits_{0}^{\infty} \intdbegin{D} D^{m}\,p(D,\tau)
\text{.}
\end{equation}
It should be noted that the first moment for large $\tau$ is known as the mean diffusion coefficient, which is obtained by a well-defined integration. This is in contrast to msd measurements, where the mean diffusion coefficient is determined by a numerical fit to the slope of the msd. By inserting \Eqref{eqn:definition of distribution of diffusivities in stationary case} into \Eqref{eqn:moments of dod} the integration over $D$ yields as a result the moments
\begin{eqnarray}
% \momentm &=& \int\limits_{0}^{\infty} \intdbegin{D} D^{m} \int \intdbegin{\disv} \delta\left(D - \frac{\disv^2}{2 N \tau}\right) p(\disv,\tau) \nonumber\\
 \momentm(\tau) &=& \frac{1}{(2 N \tau)^m}\int \intdbegin{^{N}\disv} \disv^{2m} p(\disv,\tau) \nonumber\\
  &=& (2 N \tau)^{-m} \avg{\disv^{2m}}
\label{eqn:moments of dod related to moments of propagator}
\text{.}
\end{eqnarray}
They are directly related to the moments of the distribution of displacements and, thus, to the moments of the propagator $p(\posv, t| \posv' , t')$.

\section{\label{sec:properties}Properties of the distribution of diffusivities for homogeneous anisotropic Brownian motion}

\subsection{\label{sec:distribution}Distribution of diffusivities}
For homogeneous anisotropic diffusion in $N$ dimensions, where $p(\disv,\tau)$ is a Gaussian distribution with zero mean given by \Eqref{eqn:distribution of displacements}, the computation of the distribution of diffusivities, its moments, or its characteristic function is simplified by reformulating the integral of \Eqref{eqn:definition of distribution of diffusivities in stationary case}. Applying the coordinate transformation $\disv = \ten{Q} \vektor{q}$ with $\ten{Q} = \sqrt{2 \tau} \ten{O}\transpose\sqrt{\tensorDiagD}$ gives for the distribution of diffusivities
\begin{align}
    \label{eqn:transformation of dod}
	\paniso{N}(D) = & \int \intdbegin{q_1} \dotsi \int \intdbegin{q_N}  \nonumber\\
	& \times \delta\left(D - \frac{1}{N}\sum_{i=1}^{N} D_i q_i^2\right) \prod_{j=1}^{N} p_{(0,1)}(q_j)
\text{,}
\end{align}
where $p_{(0,1)}(q_j) = \frac{1}{\sqrt{2 \pi}} \exp(-\onehalf q_j^2)$. Thus, the distribution of diffusivities is calculated by integration over independent standard normally distributed variables with zero mean and unit variance. Since the msd for homogeneous anisotropic diffusion again grows linearly as in the homogeneous isotropic case, the $\tau$ dependency in the distribution of diffusivities vanishes.

By obtaining the distribution of diffusivities, for instance, from displacements along a single trajectory, information about the orientation of the diffusion tensor is lost. However, all directions contribute to the distribution and, thus, it still contains information about the diffusion coefficients corresponding to the principal axes, i.e., the eigenvalues of $\ten{D}$.

\subsection{\label{sec:characteristic function}Characteristic function, cumulants and moments}
With the transformation \Eqref{eqn:transformation of dod}, the moments and the characteristic function of the distribution of diffusivities of anisotropic Brownian motion can be calculated. For the moments, given by \Eqref{eqn:moments of dod}, this yields
\begin{equation}
 \label{eqn:general moments of dod}
 \momentm^{N\text{d}} = \frac{1}{N^m} \int \intdbegin{q_1} \dotsi \int \intdbegin{q_N} \left(\sum_{i=1}^{N} D_i q_i^2\right)^m \prod_{j=1}^{N} p_{(0,1)}(q_j)
\text{.}
\end{equation}
So, for instance, the first moment of the distribution of diffusivities is given by
\begin{equation}
  \label{eqn:general first moment of dod}
  \moment{1}^{N\text{d}} = \frac{1}{N} \sum_{i=1}^{N} D_{i} = \avg{D(\tau)}
\text{,}
\end{equation}
which is simply the arithmetic mean of all the diffusion coefficients $D_{i}$ and coincides with the slope of the msd. For higher moments of the distribution of diffusivities, it is easier to calculate its characteristic function $\Ganiso{N}(k) = \avg{\exp(\imag k D)} = \int_{0}^{\infty} \intdbegin{D} \exp(\imag k D) \paniso{N}(D)$ by substituting $\paniso{N}(D)$ from \Eqref{eqn:transformation of dod} and performing the Fourier transform to obtain
\begin{align}
    \label{eqn:general characteristic function}
		\Ganiso{N}(k) & = \prod_{j=1}^{N} \int \intdbegin{q_j} \exp\left(\imag k \frac{D_j q_j^2}{N}\right) p_{(0,1)}(q_j) \nonumber\\
        & =  \prod_{j=1}^{N} \left(1- \imag k \frac{2 D_j}{N}\right)^{-\onehalf}
\text{.}
\end{align}

From the characteristic function \Eqref{eqn:general characteristic function} the cumulants of the distribution $\paniso{N}(D)$ are obtained as
\begin{equation}
    \label{eqn:general cumulants}
    \kappa_m = \left.\frac{1}{\imag^m} \frac{\partial^m \ln \Ganiso{N}(k)}{\partial k^m}\right|_{k=0}
             = \frac{2^{m-1}(m-1)!}{N^m} \sum_{i=1}^N D_i^m
\end{equation}
for $m>0$. The moments are recursively related to the cumulants by 
\begin{equation}
    \label{eqn:relation_cumulants_moments}
    \momentm = \sum_{k=0}^{m-1}\binom{m-1}{k} \kappa_{m-k} \moment{k}
\end{equation}
with initial value $\moment{0} = 1$ \cite{Smith1995}.

It should be noted that the characteristic function in \Eqref{eqn:general characteristic function} is a product of different characteristic functions in Fourier space. Hence, the distribution of diffusivities of an $N$-dimensional anisotropic system is determined by inverse Fourier transform of the characteristic function $\paniso{N}(D) = \InverseFourierTransform{\Ganiso{N}(k)} = \InverseFourierTransform{\prod_{i=1}^{N} \Giso{1}{D_i/N}(k)}$, where $\Giso{1}{D_i/N}(k) = \FourierTransform{\piso{1}{D_i/N}(D)}$ is the Fourier transform of the one-dimensional distribution of diffusivities $\piso{1}{D_i/N}(D) = 1/\sqrt{2\pi D D_i/N} \exp\left(-N D/(2 D_i)\right)$ with diffusion coefficient $D_i/N$. Correspondingly, the distribution of diffusivities of an $N$-dimensional anisotropic system is obtained by convolution of $N$ one-dimensional distributions of diffusivities
\begin{align}
  \label{eqn:dod_via_convolution}
  \paniso{N}(D) = & \{\piso{1}{D_1/N} \ast \piso{1}{D_2/N} \ast \dotsm \ast \piso{1}{D_N/N}\}(D) \nonumber\\
  = & \int\limits_0^\infty \intdbegin{\Delta_1} \dotsi \int\limits_0^\infty \intdbegin{\Delta_N} \nonumber\\
  & \times \delta\left(D - \sum_{i=1}^{N}\Delta_i\right) \prod_{j=1}^{N} \piso{1}{D_j/N}(\Delta_j)
\text{,}
\end{align}
which follows directly from \Eqref{eqn:transformation of dod}. Thus, with \EqsCommaAndref{eqn:transformation of dod}{eqn:general characteristic function}{eqn:dod_via_convolution}, we provide three equivalent expressions to determine the distribution of diffusivities in terms of the eigenvalues $D_i$ of $\ten{D}$. Depending on the considered experimental system each representation offers its own advantages.

\subsection{\label{sec:asymptotic decay}Asymptotic decay}
In the following, we present the asymptotic behavior of the distribution of diffusivities for homogeneous anisotropic Brownian motion. We show how the anisotropy of the process can be identified.

Considering an $\ParamDegeneracy$-fold degeneracy of the largest diffusion coefficient with $D_1 = \dotsb = D_\ParamDegeneracy > D_{\ParamDegeneracy+1} \geq \dotsb \geq D_{N}$ the distribution of diffusivities of the homogeneous anisotropic system is obtained from the convolution
\begin{equation}
 \label{eqn:dod from convolution for degeneracy of largest diffusion coefficient}
 \paniso{N}(D) = \{\piso{\ParamDegeneracy}{D_1/N} \ast \piso{1}{D_{(\ParamDegeneracy+1)}/N} \ast \dotsm \ast \piso{1}{D_N/N}\}(D)
\text{,}
\end{equation}
where $\piso{\ParamDegeneracy}{D_1/N}(D)$ is the distribution of diffusivities of the $\ParamDegeneracy$-dimensional isotropic system \Eqref{eqn:dod homogeneous isotropic system in n dimensions} with diffusion coefficient $D_c = D_1/N$, which results from the convolution of $\ParamDegeneracy$ identical one-dimensional distributions $\piso{1}{D_1/N}(D)$.

For $D \gg D_1 D_{\ParamDegeneracy+1}/(D_1-D_{\ParamDegeneracy+1})$ an asymptotic expansion for large $D$ is performed and yields the asymptotic behavior of \Eqref{eqn:dod from convolution for degeneracy of largest diffusion coefficient}
\begin{align}
 \label{eqn:asymptotic_Nd_aniso}
 \paniso{N}(D) \overset{D \to \infty}{\sim}
  & \left(\frac{N}{2 D_1}\right)^{\frac{\ParamDegeneracy}{2}} \frac{D^{\frac{\ParamDegeneracy}{2}-1}}{\Gamma(\frac{\ParamDegeneracy}{2})} \nonumber\\
  & \times \exp\left(-\frac{N}{2 D_1} D\right) \prod_{j=\ParamDegeneracy+1}^{N} \sqrt{\frac{D_1}{D_1-D_j}}
\text{.}
\end{align}
Thus, the leading behavior in the logarithmic representation is given by
\begin{equation}
 \label{eqn:log_asymptotic_Nd_aniso}
 \log \paniso{N}(D) \overset{D \to \infty}{\sim} -\frac{N}{2 \asymptoticD} D
\text{,}
\end{equation}
with $\asymptoticD = \max(D_1, D_2, \dotsc, D_N)$, i.e., an exponential decay involving the largest diffusion coefficient of the anisotropic system.

In homogeneous isotropic systems $\asymptoticD$, which describes the asymptotic decay, is equal to the isotropic diffusion coefficient $D_c$, which further coincides with the first moment $\meanD$. The corresponding distribution of diffusivities is a $\chi^2$-distribution given by \Eqref{eqn:dod homogeneous isotropic system in n dimensions}. This is in contrast to the anisotropic case, where $\meanD < \asymptoticD$. Thus, a discrepancy between $\meanD$ and $\asymptoticD$ leads to deviations from the $\chi^2$-distribution and rules out a homogeneous isotropic process. In general, this can be exploited to detect that the observed system comprises more than one diffusion coefficient. By further assuming homogeneity such a system is identified as an anisotropic one.

A quantitative measure for the discrepancy between $\meanD$ and $\asymptoticD$ is given by
\begin{equation}
  \label{eqn:anisotropy measure from asymptotic decays}
  \anisotropyMeasure = \frac{\asymptoticD}{\meanD} - 1
\end{equation}
which characterizes the deviation from the homogeneous isotropic case. Thus, for homogeneous systems it quantifies the anisotropy of the process. In cases where both values coincide, i.e., the system is isotropic, $\anisotropyMeasure$ becomes zero. In contrast, if one diffusion coefficient is much larger than all others, $\avg{D} \to \asymptoticD/N$ resulting in $\anisotropyMeasure = N-1$, which denotes the largest possible anisotropy in $N$ dimensions. Thus, $\anisotropyMeasure$ is a measure of the anisotropy, but it is not suitable to compare systems of different dimensionality $N$. It should be noted that similar measures exist \cite{Hess1991, PhysRevE.75.021112}.

From experimental data, both quantities for the anisotropy measure \Eqref{eqn:anisotropy measure from asymptotic decays} can be determined easily. The mean diffusion coefficient $\meanD$ corresponds to the first moment of the distribution of diffusivities and is obtained by averaging the diffusivities. The decay for large $D$ is obtained from a fit to $f(D) = c \exp(-\lambda_\text{fit} D)$ to calculate $\asymptoticD= N/(2 \lambda_\text{fit})$. The actual dimensionality $\effectiveDimensionality$ of processes observed in $N \geq \effectiveDimensionality$ dimensions can be estimated with $\effectiveDimensionality = 2 \meanD \lambda_\text{fit}$ leading to $\anisotropyMeasure = N/\effectiveDimensionality -1$. For example, if an observed $N$-dimensional motion yields the largest anisotropy value of $\anisotropyMeasure = N-1$, the process is effectively a one-dimensional motion.

\subsection{\label{sec:general reconstruction}Reconstruction of the diffusion tensor}
For experiments it is of great interest to reconstruct the diffusion tensor $\ten{D}$ from measurements. If complete information about the trajectories is available, the diffusion tensor of the homogeneous anisotropic process can be estimated via the displacements. By defining tensorial diffusivities analogously to \Eqref{eqn:single-particle diffusivity}
\begin{equation}
    \label{eqn:tensorial diffusivities}
    D^{ij}_t(\tau) = \frac{[x_i(t+\tau)-x_i(t)][x_j(t+\tau)-x_j(t)]}{2 \tau}
\text{,}
\end{equation}
where $x_{i}(t)$ denotes the $i$-th component of the $N$-dimensional trajectory $\posv(t)$, the linear $\tau$ dependence of the mixed displacements is removed. These tensorial diffusivities are simply averaged
\begin{equation}
     \label{eqn:tensor estimation}
     D_{ij} = \avg{D^{ij}_t(\tau)}
\end{equation}
providing an estimator for the corresponding elements of $\ten{D}$. Here, $\avg{\dotso}$ either denotes a time average or an ensemble average depending on the available data.

\section{\label{sec:Projection}Projection to an \texorpdfstring{$M$}{M}-dimensional subspace}

Due to experimental restrictions the complete trajectory is often not accessible but its projection on an $M$-dimensional subspace can be measured. Such processes are commonly known as observed diffusion \cite{dembo1986, campillo1989}.

The projection of the distribution of displacements \Eqref{eqn:distribution of displacements} on the considered subspace is the marginal probability density
\begin{equation}
    \label{eqn:projected_distribution of displacements}
    p(\disv^M_{\vektor{\alpha}},\tau) = \int \intdbegin{r_{\alpha_{1}}} \dotsi \int \intdbegin{r_{\alpha_{N-M}}} p(\disv,\tau)
    \text{,}
\end{equation}
where $r_{\alpha_{i}}, i = 1,\dotsc,(N-M)$ denotes $(N-M)$ arbitrarily chosen directions which are integrated out. The vector $\vektor{\alpha} = \begin{pmatrix} \alpha_1 & \dotso & \alpha_{N-M}\end{pmatrix}$ contains the indices $\alpha_{i}$ describing which elements of $\vektor{r}$ are omitted. Alternatively, the projected distribution of displacements is computed by the $M$-dimensional inverse Fourier transform of the characteristic function of $p(\disv,\tau)$ where the components, $k_{\alpha_{i}}=0,\, \forall i \in \{1,\dotsc,N-M\}$, which correspond to the chosen directions, are discarded. Hence, the distribution of displacements of the subspace is
\begin{equation}
    \label{eqn:projected distribution of displacements via Fourier transform of characteristic function}
    p(\disv^M_{\vektor{\alpha}},\tau) = \int \intdbegin{^{M}\vektor{k}^M_{\vektor{\alpha}}} \frac{1}{(2\pi)^{M}}\exp\left[-\imag (\vektor{k}^M_{\vektor{\alpha}})\transpose \disv^M_{\vektor{\alpha}}\right] G(\vektor{k}^M_{\vektor{\alpha}})
\end{equation}
with the characteristic function of the projected propagator $G(\vektor{k}^M_{\vektor{\alpha}}) = \exp\bigl[- (\vektor{k}^M_{\vektor{\alpha}})\transpose \submatrixOfSigma \vektor{k}^M_{\vektor{\alpha}}\bigr]$. The vector $\vektor{k}^M_{\vektor{\alpha}}$ is an $M$-dimensional sub-vector of the complete $k$-space and $\submatrixOfSigma$ denotes a principal $M \times M$ submatrix of $\mat{\Sigma}$ obtained by deletion of rows and columns with corresponding indices $\alpha_i$.

The distribution of diffusivities of such a projected diffusion process is calculated analogously to \Eqref{eqn:definition of distribution of diffusivities in stationary case} by integrating over $\disv^M_{\vektor{\alpha}}$. Since $\mat{\Sigma} = 2 \tau \ten{D}$ is a symmetric, positive definite matrix for $\tau > 0$, all principal submatrices $\submatrixOfSigma$ are symmetric, positive definite matrices as well and can be diagonalized. Hence, the projected distribution of diffusivities has the $M$-dimensional form of the generic expression \EqsCommaOrref{eqn:transformation of dod}{eqn:general characteristic function}{eqn:dod_via_convolution}. However, it depends on the eigenvalues $D^M_{k,\vektor{\alpha}}, k = 1,\dotsc,M$ of the projected diffusion tensor $\ten{D}^M_{\vektor{\alpha}} = \submatrixOfSigma/(2 \tau)$. If the eigenvalues of $\ten{D}$ are identified as
\begin{equation}
  \label{eqn:relation of eigenvalues}
    D_1 \geq D_2 \geq \dotsb \geq D_N
\end{equation}
and the eigenvalues of $\ten{D}^{N-1}_{\vecalphaOneElement{\alpha}}$ are
\begin{equation}
  \label{eqn:relation of eigenvalues of submatrix}
    D^{N-1}_{1,\vecalphaOneElement{\alpha}} \geq D^{N-1}_{2,\vecalphaOneElement{\alpha}} \geq \dotsb \geq D^{N-1}_{N-1,\vecalphaOneElement{\alpha}},\; \forall \alpha \in \{1,\dotsc,N\} \text{,}
\end{equation}
the well-known interlacing inequalities \cite{Cauchy1829} require
\begin{equation}
    \label{eqn:cauchy interlacing inequalities}
    D_k \geq D^{N-1}_{k,\vecalphaOneElement{\alpha}} \geq D_{k+1},\; \forall k \in \{1,\dotsc,N-1\}
\end{equation}
for all $\alpha \in \{1,\dotsc,N\}$. This expression is applied recursively (N-M) times to obtain a relation for the eigenvalues of the principal $M\times M$ submatrix \cite{Fan1957}
\begin{equation}
    D_k \geq D^M_{k,\vektor{\alpha}} \geq D_{k+N-M},\; \forall k \in \{1,\dotsc,M\}
\end{equation}
for arbitrary $\vektor{\alpha}$. By implication, if at least two eigenvalues of the submatrix $\ten{D}^M_{\vektor{\alpha}}$ differ, i.e., the projected process is anisotropic, \Eqref{eqn:cauchy interlacing inequalities} states recursively that the complete process is anisotropic as well. Thus, the distribution of diffusivities of the projected $N$-dimensional anisotropic Brownian motion may already indicate the anisotropy of the complete process as well as the magnitude of one of the involved diffusion coefficients. However, a single projection is not sufficient to obtain the underlying diffusion coefficients.

Nevertheless, it is possible to estimate the bounds of the diffusion coefficients. The lower bound of the eigenvalues is given by zero, due to the positive semidefiniteness of $\ten{D}$. An upper bound for the largest eigenvalue can be found if enough projections or submatrices are available to comprise all diagonal elements of $\ten{D}$. By use of the relation between the trace of an $N\times N$ matrix $\ten{A}$ and its eigenvalues $\lambda_i$, $\tr\ten{A} = \sum_i \lambda_i$, subtotals of the trace of $\ten{D}$ are given by the sum of the eigenvalues of the respective submatrices. If the non-overlapping orthogonal projections of $\ten{D}$ defined by $\vektor{\alpha}$ compose a partition of the set $\{1,\dotsc,N\}$, the trace of the tensor is given by
\begin{equation}
    \tr\ten{D} = \sum_{i=1}^N D_i = \sum_{\vektor{\alpha}} \tr\ten{D}^M_{\vektor{\alpha}} = \sum_{\vektor{\alpha}} \sum_k D^M_{k,\vektor{\alpha}}
\end{equation}
with $\dot{\bigcup} \vektor{\alpha} = \{1,\dotsc,N\}$, where the partition elements $\vektor{\alpha}$ do not necessarily have identical dimensionality.

For example, if one measures the eigenvalues of two non-overlapping projections of a $3 \times 3$ diffusion tensor $\ten{D}$, the trace of $\ten{D}$ is given by
\begin{align}
    \tr\ten{D} & = \tr\ten{D}^1_{\vecalphaTwoElements{1}{3}} + \tr\ten{D}^2_{\vecalphaOneElement{2}} \nonumber\\
               & = D_{1,\vecalphaTwoElements{1}{3}}^1 + D_{1,\vecalphaOneElement{2}}^2 + D_{2,\vecalphaOneElement{2}}^2
\text{.}
\end{align}
Thus, the eigenvalue inequalities for that example using the relations above are given by
\begin{align}
    \tr\ten{D} \geq & D_1 \geq \max(D_{1,\vecalphaTwoElements{1}{3}}^1, D_{1,\vecalphaOneElement{2}}^2) \geq D_2 \nonumber \\
    D_2 \geq & \min(D_{1,\vecalphaTwoElements{1}{3}}^1, D_{2,\vecalphaOneElement{2}}^2) \geq D_3 \geq 0
\text{,}
\end{align}
which allows a rough estimation of the diffusion coefficients from the given projections.

\section{\label{sec:Specific systems}Specific systems}

\subsection{\label{sec:Anisotropy2d}Two-dimensional systems}
The distribution of diffusivities of a two-dimensional homogeneous anisotropic system can be calculated explicitly, for instance, via \Eqref{eqn:dod_via_convolution}, resulting in
\begin{align}
      \label{eqn:dod_2d_aniso}
      \paniso{2}(D) =& \int\limits_0^\infty \intd{\Delta_1} \int\limits_0^\infty \intdbegin{\Delta_2}  \delta\left[D-(\Delta_1+\Delta_2)\right] \nonumber\\
	  &\times \piso{1}{D_1/2}(\Delta_1) \piso{1}{D_2/2}(\Delta_2) \nonumber\\
      =& \frac{\exp\left[-\onehalf \left(\frac{1}{D_1}+\frac{1}{D_2}\right) D\right]}{\sqrt{D_1D_2}} \modifiedBessel{0}\left[\onehalf \left(\frac{1}{D_1}-\frac{1}{D_2}\right) D \right]
\end{align}
where $\modifiedBessel{0}(x)$ denotes the modified Bessel function of the first kind. The first two moments of this distribution, as given by \Eqref{eqn:moments of dod}, yield
\begin{align}
  \label{eqn:first moment of 2d anisotropic system}
  \avg{D} = \moment{1} = \frac{1}{2}(D_1+D_2)
\end{align}
and
\begin{align}
  \label{eqn:second moment of 2d anisotropic system}
  \avg{D^2} = \moment{2} = \frac{1}{4}(3 D_1^2+2D_1D_2+3D_2^2)
  \text{.}
\end{align}
Hence, the mean diffusion coefficient coincides with the arithmetic mean of the diffusion coefficients belonging to the two directions of the anisotropic system as expected from \Eqref{eqn:general first moment of dod}. Solving the simultaneous \EqsAndref{eqn:first moment of 2d anisotropic system}{eqn:second moment of 2d anisotropic system} yields the expression
\begin{equation}
  \label{eqn:eigenvalues2d}
  D_{1,2} = \moment{1} \pm \sqrt{\moment{2} - 2 \moment{1}^2}
\end{equation}
to obtain the diffusion coefficients $D_1$ and $D_2$ from the moments.

\begin{figure}[ht]
	\includegraphics{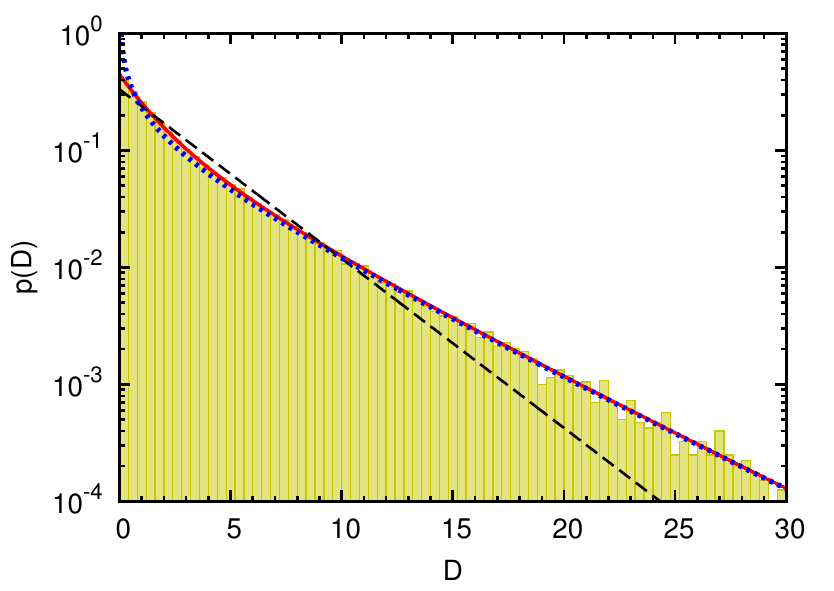}
\caption{\label{fig:dod of 2d anisotropic diffusion}The distribution of diffusivities (histogram) from a simulated trajectory of a homogeneous anisotropic diffusion process in two dimensions with diffusion tensor $\ten{D}$ given by \Eqref{eqn:simulation_tensor_2d} agrees well with the analytic distribution of diffusivities (solid line) from \Eqref{eqn:dod_2d_aniso} with $D_1 = 5$ and $D_2 = 1$ denoting the eigenvalues of $\ten{D}$. Additionally, the asymptotic function \Eqref{eqn:asymptotic_2d_aniso} (dotted line, $\asymptoticD=5$) agrees reasonably for large $D$. Furthermore, a distribution of diffusivities (dashed line) of two-dimensional isotropic diffusion with the same mean diffusion coefficient $D_c = \avg{D} = (D_1+ D_2)/2 = 3$ is shown for comparison. The different asymptotic decays are clearly visible and allow the distinction from homogeneous isotropic processes.}
\end{figure}

The asymptotic behavior of \Eqref{eqn:dod_2d_aniso} for large $D$ is given by \Eqref{eqn:asymptotic_Nd_aniso} and yields
\begin{equation}
    \label{eqn:asymptotic_2d_aniso}
    \paniso{2}(D) \overset{D \to \infty}{\sim} \frac{\exp\left(-\frac{D}{\asymptoticD}\right)}{\sqrt{\abs{D_1-D_2} \pi D}}
\end{equation}
with $\asymptoticD = \max(D_1,D_2)$. Thus, the asymptotic behavior in the logarithmic representation is given by
\begin{equation}
    \label{eqn:log_asymptotic_2d_aniso}
	\log \paniso{2}(D) \overset{D \to \infty}{\sim} -\frac{D}{\asymptoticD}
\text{,}
\end{equation}
which corresponds to the decay of the distribution of diffusivities in two-dimensional homogeneous isotropic systems with diffusion coefficient $\asymptoticD$, i.e., an exponential decay with the largest diffusion coefficient of the anisotropic system. Accordingly, the smallest diffusion coefficient is given by $2\meanD-\asymptoticD$. From the asymptotic decay and the mean diffusion coefficient the anisotropy of the system is characterized by \Eqref{eqn:anisotropy measure from asymptotic decays} and corresponds to the ratio
\begin{equation}
 \label{eqn:ratio characterizing anisotropy}
 \anisotropyMeasure = \frac{\abs{D_1-D_2}}{D_1+D_2} = \frac{\sqrt{\moment{2} - 2 \moment{1}^2}}{\moment{1}}
\text{,}
\end{equation}
which is also related to the moments.

The diffusion coefficients $D_1, D_2$ can also be obtained from the asymptotic behavior for vanishing $D$. Since
\begin{equation}
    \label{eqn:asymptotic behavior for D to 0}
    \lim_{D \to 0} \paniso{2}(D) = (D_1 D_2)^{-\onehalf}
\text{,}
\end{equation}
the corresponding value in experimental data is determined by extrapolating the distribution of diffusivities in a log-log plot towards $D = 0$. In conjunction with an estimate of the largest diffusion coefficient from a fit to \Eqref{eqn:log_asymptotic_2d_aniso} both diffusion coefficients can be identified. This provides a consistency check for the calculation via the moments of the distribution of diffusivities given in \Eqref{eqn:ratio characterizing anisotropy}.

To substantiate our analytical expressions by results from simulations, a random walk was performed by numerical integration of the Langevin equation, \Eqref{eqn:langevin-nd}, in two dimensions using the diffusion tensor
\begin{equation}
    \label{eqn:simulation_tensor_2d}
    \ten{D} = \left(\begin{array}{cc} 4 & \sqrt{3} \\ \sqrt{3} & 2\end{array}\right)
\end{equation}
with eigenvalues $D_{1} = 5$ and $D_{2} = 1$. The obtained trajectory of the two-dimensional homogeneous anisotropic diffusion process consisted of $10^5$ displacements and its distribution of diffusivities is depicted in \Figref{fig:dod of 2d anisotropic diffusion}. The agreement of the normalized histogram from simulated data with the analytic distribution \Eqref{eqn:dod_2d_aniso} is obvious. Deviations between simulation and the analytic curve for large $D$ are due to insufficient statistics from the finite number of displacements. Moreover, \Figref{fig:dod of 2d anisotropic diffusion} shows the mono-exponential behavior corresponding to isotropic diffusion in two dimensions for comparison. Although the mean diffusion coefficients of both processes coincide, the asymptotic decays of the distributions differ. The reason is the asymptotic behavior given by \Eqref{eqn:asymptotic_2d_aniso} in the anisotropic case which decays exponentially with the largest eigenvalue for large $D$ as depicted in the figure. In contrast, for the isotropic system the asymptotic decay corresponds to the mean diffusion coefficient resulting in the observed quantitative difference. Furthermore, the distributions are qualitatively different for small $D$. A characteristic difference between isotropic and anisotropic systems is the convex shape in the logarithmic representation of the anisotropic distribution of diffusivities. This intuitively results from the two different exponential decays related to the distinct diffusion coefficients $D_1$ and $D_2$. In a more rigorous way, since $\frac{\dd^2}{\dd D^2}\log \paniso{2}(D) \geq 0$, with the equal sign being valid only for isotropic diffusion, the anisotropic distribution of diffusivities is a superconvex function \cite{kingman1961}.

For experimental data, it is easy to calculate the first two moments $\moment{1}$  and $\moment{2}$ by averaging the short-time diffusivities of \Eqref{eqn:single-particle diffusivity} and their squares, respectively. The averaging is accomplished either along a single trajectory or from an ensemble of trajectories avoiding any numerical fit. The first two moments are sufficient to calculate $D_{1}$ and $D_{2}$ by \Eqref{eqn:eigenvalues2d}.

For the sample trajectory used in \Figref{fig:dod of 2d anisotropic diffusion} the first two moments are determined to be $\measured{\moment{1}} = 2.987$ and $\measured{\moment{2}} = 21.72$. According to \Eqref{eqn:eigenvalues2d}, the underlying diffusion coefficients yield $\measured{D}_{1} = 4.956$ and $\measured{D}_{2} = 1.018$. These values agree well with the eigenvalues of the tensor \Eqref{eqn:simulation_tensor_2d}, which was used as input parameter of the simulation. The resulting value of $\anisotropyMeasure=2/3$ indicates a considerable anisotropy of the process.

\subsubsection{\label{sec:limits Anisotropy2d}Limiting cases}
In the case of identical diffusion coefficients for both directions the anisotropy vanishes as discussed for \Eqref{eqn:ratio characterizing anisotropy}. The resulting isotropic diffusion process is characterized by a single diffusion coefficient $D_c = D_1 = D_2$. Hence, \Eqref{eqn:dod_2d_aniso} simplifies to the well-known distribution of single-particle diffusivities of two-dimensional isotropic diffusion \cite{bauer2011}
\begin{equation}
 \label{eqn:homogeneous dod in 2d}
 \piso{2}{D_c}(D) = \frac{\exp\left(-\frac{D}{D_c}\right)}{D_c}
\end{equation}
given by an exponential function.

If, on the contrary, the anisotropy is large, diffusion in one direction will be suppressed. Without loss of generality, this is accomplished by sending one of the diffusion coefficients to zero. Thus, by taking the limit of vanishing $D_2$, the distribution of diffusivities \Eqref{eqn:dod_2d_aniso} is simplified to
\begin{equation}
 \label{eqn:homogeneous dod equal to 1d with half diffusion coefficient}
 \piso{1}{D_1}(D) = \lim_{D_2 \to 0} \paniso{2}(D) = \frac{\exp\left(-\frac{D}{D_1}\right)}{\sqrt{\pi D_1 D}}
\text{,}
\end{equation}
which has the structure of the distribution of diffusivities of one-dimensional diffusion \cite{bauer2011}. Since diffusion into the perpendicular direction is prohibited, as expected, it qualitatively leads to the observation of a one-dimensional process. This can be identified by the characteristic factor $D^{-1/2}$ due to which the distribution of diffusivities diverges for small $D$. Applying \Eqref{eqn:moments of dod} the first moment of \Eqref{eqn:homogeneous dod equal to 1d with half diffusion coefficient}, i.e., the mean diffusion coefficient, yields $\meanD = D_1/2$. The factor of $1/2$ results from the single-particle diffusivities \Eqref{eqn:single-particle diffusivity} with $N=2$ assuming that a two-dimensional process is observed. However, due to the suppression of one direction this assumption is no longer valid and $N=1$ should have been used instead. This conclusion is also obtained from the anisotropy value $\anisotropyMeasure = 1$, which is equal to its maximum value for two-dimensional anisotropic processes since effectively one-dimensional motion is observed.

\subsubsection{\label{sec:projected2dto1d}Reconstruction of \texorpdfstring{$\ten{D}$}{D} }
In addition to the eigenvalues, it is sometimes of interest to determine the orientation of the principal axes of the system relative to the given frame of reference. This is achieved by the reconstruction of the diffusion tensor
\begin{equation}
    \ten{D} = \begin{pmatrix} D_{11} & D_{12} \\ D_{12} & D_{22}\end{pmatrix}
\text{,}
\end{equation}
where the off-diagonal elements are labeled identically due to symmetry reasons. The reconstruction is accomplished in two ways either by considering the complete two-dimensional trajectory or by using one-dimensional projections of the trajectory.

In the first approach the tensorial diffusivities of \Eqref{eqn:tensorial diffusivities} are used to obtain the tensor entries of $\ten{D}$. In accordance with \Eqref{eqn:tensor estimation} the tensor elements are estimated by averaging the tensorial diffusivities along a trajectory or over an ensemble. Moreover, the eigenvalues of the tensor $\ten{D}$ are expressed by its entries
\begin{equation}
    \label{eqn:eigenvalues from entries 2d}
    D_{1,2} = \onehalf\left(D_{11} + D_{22} \pm \sqrt{(D_{11}-D_{22})^2 + 4 D_{12}^2}\right)
\end{equation}
and correspond to the diffusion coefficients of the system.

For the sample trajectory used in \Figref{fig:dod of 2d anisotropic diffusion} the measured values $\measured{D}_{ij}$ yield the diffusion tensor
\begin{equation}
    \measured{\ten{D}} = \begin{pmatrix} 3.983 & 1.719 \\ 1.719 & 1.990 \end{pmatrix}
\text{,}
\end{equation}
which agrees reasonably with the input parameters of the simulation. The eigenvalues from this measured tensor $\measured{D}_{1} = 4.973$ and $\measured{D}_{2} = 1.000$ show a good agreement with the exact eigenvalues of the input tensor $D_1 = 5$ and $D_2 = 1$.

The second approach determines the tensor $\ten{D}$ exclusively from one-dimensional projections of the trajectory. In order to obtain results, at least three different projections are necessary. For simplicity, it is preferable to use projections along two perpendicular axes, which define the frame of reference for $\ten{D}$. Furthermore, a projection onto an axis is required which is rotated about an angle $\theta$ relatively to the frame of reference. In such a setup, the first moments of the distribution of diffusivities related to the first two projections are identical to the averaged tensorial diffusivities $\avg{D_t^{11}(\tau)}$ and $\avg{D_t^{22}(\tau)}$. Thus, they yield the two diagonal elements of $\ten{D}$. The first moment of the third projection measures the leading diagonal element $D^\theta_{11}$ of the rotated tensor $\ten{D}^\theta = \ten{R}(\theta)\transpose\ten{D}\ten{R}(\theta)$ with rotation tensor
$\ten{R}(\theta) = \begin{pmatrix} \cos \theta & -\sin \theta \\ \sin \theta & \cos \theta\end{pmatrix}$. This additional value is sufficient to obtain the off-diagonal element of $\ten{D}$ from
\begin{equation}
    D_{12} = \frac{D^\theta_{11} - D_{11} \cos^2 \theta  - D_{22} \sin^2 \theta}{\sin(2 \theta)}
\text{.}
\end{equation}
For the calculation, any projection of the trajectory onto an arbitrary one-dimensional axis, i.e., any $\theta$, can be used except directions perpendicular or parallel to axes of the frame of reference, i.e., angles $\theta$ which are multiples of $\pi/2$. It should be emphasized that the reconstruction from the distribution of diffusivities of projected trajectories is possible although the definition of the diffusivities omit any directional information.

In the example with $\measured{D}_{11} = 3.983$, $\measured{D}_{22} = 1.990$ and a measured $\measured{D}_{11}^{5\pi/12} = 2.983$, the off-diagonal element yields $D_{12} = 1.719$, which is in good agreement with the value $\sqrt{3} \approx 1.732$ appearing as input parameter of the simulation.

In conclusion, it depends on the constraints of the experiment which of both approaches is more practicable. In either way the complete diffusion tensor $\ten{D}$ is reconstructed reasonably well.

\subsection{\label{sec:Anisotropy3d}Three-dimensional systems}
Analogous to the two-dimensional case, it is possible to calculate the distribution of diffusivities for three-dimensional systems either by inverse Fourier transform of the general characteristic function \Eqref{eqn:general characteristic function} or by the convolution \Eqref{eqn:dod_via_convolution}. In both cases the analytical integration cannot be performed completely. However, the integration can be accomplished numerically. By integrating two variables \Eqref{eqn:dod_via_convolution} is reduced to
\begin{align}
      \label{eqn:dod_3d_aniso}
      \paniso{3}(D) &= \int\limits_0^\infty \intd{\Delta_1}\! \int\limits_0^\infty \intd{\Delta_2} \int\limits_0^\infty \intdbegin{\Delta_3} \delta\left[D-(\Delta_1+\Delta_2+\Delta_3)\right] \nonumber\\
      &\times \piso{1}{D_1/3}(\Delta_1) \piso{1}{D_2/3}(\Delta_2) \piso{1}{D_3/3}(\Delta_3)  \nonumber\\
      &= \int\limits_0^{D} \intdbegin{\Delta_1} \left(\frac{3}{2}\right)^{3/2}\frac{1}{\sqrt{\pi D_1 D_2 D_3 \Delta_1}} \nonumber\\
      &\times\exp\left\{-\frac{3}{4}\left[\left(\frac{1}{D_2}+\frac{1}{D_3}\right)(D-\Delta_1) + \frac{2\Delta_1}{D_1}\right]\right\}\nonumber\\
      &\times \modifiedBessel{0}\left[\frac{3}{4}\left(\frac{1}{D_3} - \frac{1}{D_2}\right)(D-\Delta_1)\right]
\text{.}
\end{align}
For further simplification, a series expansion of the modified Bessel function $\modifiedBessel{0}(x)$ can be applied, which allows performing the last integration. However, this only results in a converging sum, which cannot be simplified any further.

By using the general expression of the cumulants \Eqref{eqn:general cumulants} and the relation between cumulants and moments \Eqref{eqn:relation_cumulants_moments}, the first three moments of the distribution of diffusivities of three-dimensional homogeneous anisotropic diffusion processes are
\begin{equation}
    \label{eqn:m1_general_3d}
    \moment{1} = \frac{1}{3}(D_1 + D_2 + D_3)
\text{,}
\end{equation}
\begin{equation}
    \label{eqn:m2_general_3d}
    \moment{2} = \frac{1}{9}\left[(D_1 + D_2 + D_3)^2 + 2(D_1^2 + D_2^2 + D_3^2)\right]
\text{,}
\end{equation}
and
\begin{align}
    \label{eqn:m3_general_3d}
    \moment{3} & = \frac{1}{9}\left[5 D_1^3 + 3 D_1^2 (D_2 + D_3)\right.\nonumber \\
               & + D_1 (3 D_2^2 + 2 D_2 D_3 + 3 D_3^2)\nonumber \\
               & \left.+ (D_2 + D_3) (5 D_2^2 - 2 D_2 D_3 + 5 D_3^2)\right]
    \text{.}
\end{align}
These expressions are similar to \EqsToref{eqn:first moment of 2d anisotropic system}{eqn:second moment of 2d anisotropic system} and relate the moments of the distribution of diffusivities to diffusion coefficients $D_1$ to $D_3$ of the anisotropic process. By solving simultaneously \EqsToref{eqn:m1_general_3d}{eqn:m3_general_3d}, the underlying diffusion coefficients are determined by the measured moments of the distribution. The solution comprises six triplets ($D_1$ to $D_3$), which are permutations of the three diffusion coefficients. Due to the cubic contributions in \Eqref{eqn:m3_general_3d} the expressions are too lengthy to be shown here but can be easily obtained.

\begin{figure}[ht]
	\includegraphics{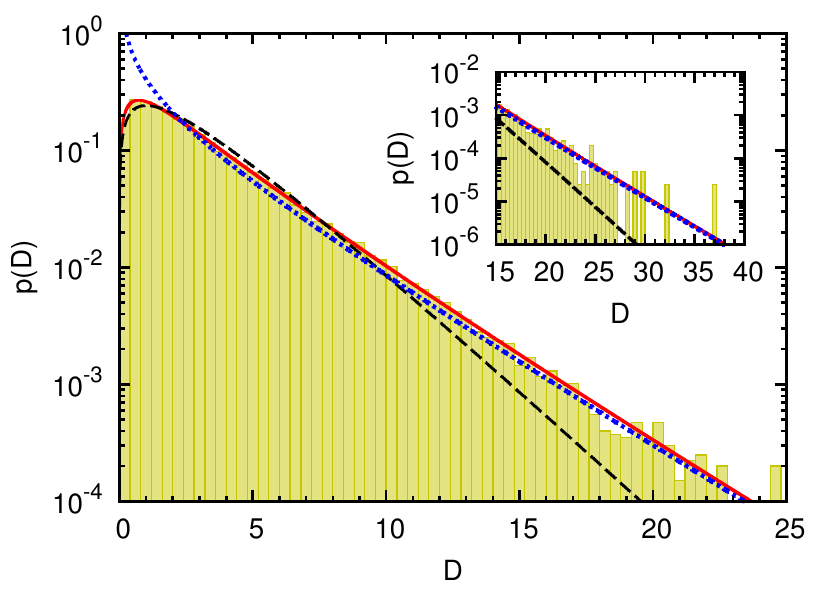}
\caption{\label{fig:dod of 3d_biaxial anisotropic diffusion}The distribution of diffusivities (histogram) from one simulated trajectory of a homogeneous anisotropic diffusion process in three dimensions with diffusion tensor $\ten{D}$ given by \Eqref{eqn:simulation_tensor_3d} agrees well with the distribution of diffusivities (solid line) obtained from numerical integration of \Eqref{eqn:dod_3d_aniso}, using the eigenvalues $D_1 = 5$, $D_2 = 3$ and $D_3 = 1$ of tensor $\ten{D}$. For comparison, the distribution of diffusivities (dashed line) of an isotropic diffusion process in three dimensions, given by \Eqref{eqn:dod 3d isotropic homogeneous}, is shown, where the same mean diffusion coefficient $D_c = \avg{D} = (D_1+ D_2 + D_3)/3 = 3$ as in the anisotropic process was used. The different asymptotic decays are clearly visible and allow the distinction from homogeneous isotropic processes. In the inset, the asymptotic function \Eqref{eqn:asymptotic_3d_aniso} (dotted line) agrees reasonably for large $D$.}
\end{figure}

The asymptotic behavior of \Eqref{eqn:dod_3d_aniso} for large $D$ is given by \Eqref{eqn:asymptotic_Nd_aniso}, which assumes $D_1 > D_2 > D_3$, and results in
\begin{equation}
    \label{eqn:asymptotic_3d_aniso}
    \paniso{3}(D) \overset{D \to \infty}{\sim} \frac{\sqrt{3 D_1}\exp\left(-\frac{3 D}{2 D_1}\right)}{\sqrt{2 \pi (D_1-D_2)(D_1-D_3) D}}
\text{.}
\end{equation}
Thus, the behavior in the logarithmic representation is determined by
\begin{equation}
    \label{eqn:log_asymptotic_3d_aniso}
	\log \paniso{3}(D) \overset{D \to \infty}{\sim} -\frac{3 D}{2 \asymptoticD}
\text{,}
\end{equation}
which corresponds to the asymptotic decay of a three-dimensional isotropic distribution of diffusivities with $\asymptoticD = \max(D_1,D_2,D_3)$. The anisotropy measure \Eqref{eqn:anisotropy measure from asymptotic decays} in the three-dimensional case corresponds to
\begin{equation}
  \anisotropyMeasure = \frac{(\asymptoticD-D_1) + (\asymptoticD-D_2) + (\asymptoticD-D_3)}{D_1+D_2+D_3}
\text{,}
\end{equation}
which considers the differences of the individual diffusion coefficients to characterize the anisotropy. It is obvious that the largest anisotropy yields $\anisotropyMeasure=2$.

In order to substantiate our results by simulated data, the simulation of a three-dimensional homogeneous anisotropic random walk was performed using the diffusion tensor
\begin{equation}
    \label{eqn:simulation_tensor_3d}
    \ten{D} = \begin{pmatrix}
	  4 & -\frac{\sqrt{3}}{2} & -\frac{1}{2} \\
	  -\frac{\sqrt{3}}{2} & \frac{13}{4} & \frac{3\sqrt{3}}{4} \\
	  -\frac{1}{2} & \frac{3\sqrt{3}}{4} & \frac{7}{4}
	\end{pmatrix}
\text{.}
\end{equation}
The obtained trajectory consists of $10^5$ displacements and its distribution of diffusivities is depicted in \Figref{fig:dod of 3d_biaxial anisotropic diffusion}. The distribution of diffusivities from the simulated trajectory shows a good agreement with the curve obtained from numerical integration of \Eqref{eqn:dod_3d_aniso}. The deviations for larger values of $D$ result from the finite simulation time, i.e., its insufficient statistics. Furthermore, \Figref{fig:dod of 3d_biaxial anisotropic diffusion} shows the distribution of an isotropic system where a qualitative distinction at the crossover from the maximum peak to the exponential decay becomes apparent. This behavior of the curvature in the logarithmic representation depends on the observed system and is discussed in \Secref{sec:curvature}. The deviating asymptotic decay of the anisotropic process is clearly visible in \Figref{fig:dod of 3d_biaxial anisotropic diffusion} and allows the distinction from homogeneous isotropic processes. Thus, in conjunction with the mean diffusivity the asymptotic decay provides a measure of the anisotropy. In addition, the asymptotic behavior given by \Eqref{eqn:asymptotic_3d_aniso} is depicted and provides a reasonable approximation for large $D$. The eigenvalues of $\ten{D}$ for experimental data are easily determined by measuring the leading moments of the diffusivities. For the sample trajectory used in \Figref{fig:dod of 3d_biaxial anisotropic diffusion} the first three moments result in $\measured{\moment{1}} = 2.995$, $\measured{\moment{2}} = 16.68$ and $\measured{\moment{3}} = 140.2$. By solving the simultaneous \EqsToref{eqn:m1_general_3d}{eqn:m3_general_3d}, the underlying diffusion coefficients are obtained as $\measured{D}_{1} = 4.884$, $\measured{D}_{2} = 3.153$ and $\measured{D}_{3} = 0.948$. These values agree reasonably well with the eigenvalues of the tensor \Eqref{eqn:simulation_tensor_3d}, which was used as input parameter of the simulation. The value of $\anisotropyMeasure=2/3$ indicates a considerable anisotropy of the process.

\subsubsection{\label{sec:limits Anisotropy3d}Limiting cases}
If the diffusion coefficients of all three directions coincide with $D_c = D_1 = D_2 = D_3$, the distribution of diffusivities for the three-dimensional isotropic system \cite{bauer2011}
\begin{equation}
    \label{eqn:dod 3d isotropic homogeneous}
    \piso{3}{D_c}(D) = 3\sqrt{\frac{3}{2 \pi} \frac{D}{D_c^3}} \exp\left(-\frac{3 D}{2 D_c}\right)
\end{equation}
will be obtained from \Eqref{eqn:dod_3d_aniso} in agreement with \Eqref{eqn:dod homogeneous isotropic system in n dimensions}.

If exactly two diffusion coefficients coincide, one usually refers to diffusion processes of uniaxial molecules \cite{deGennes_B95}. In this case, the general distribution of diffusivities of three-dimensional homogeneous anisotropic diffusion \Eqref{eqn:dod_3d_aniso} simplifies to
\begin{equation}
    \label{eqn:dod_3d_uniaxial}
    \puni{3}(D) = \frac{3}{2} \frac{\exp\left(-\frac{3 D}{2 \multipleEigenvalueOfD{2}}\right) \erf\left(\sqrt{\frac{3}{2}(\frac{1}{\multipleEigenvalueOfD{1}}-\frac{1}{\multipleEigenvalueOfD{2}})D}\right) }{\sqrt{\multipleEigenvalueOfD{2}(\multipleEigenvalueOfD{2}-\multipleEigenvalueOfD{1})}}
\text{,}
\end{equation}
where $\multipleEigenvalueOfD{1}$ and $\multipleEigenvalueOfD{2}$ are the eigenvalues of $\ten{D}$ with multiplicity one and two, respectively. In general, a distinction between the oblate case ($\multipleEigenvalueOfD{2} > \multipleEigenvalueOfD{1}$, disc) and the prolate case ($\multipleEigenvalueOfD{2} < \multipleEigenvalueOfD{1}$, rod) is made for uniaxial molecules. In the prolate case both square roots in \Eqref{eqn:dod_3d_uniaxial} yield complex numbers. However, with $\erf(\sqrt{-x})/\sqrt{-y} = \erfi(\sqrt{x})/\sqrt{y}$ for $x,y >0$ and $x,y \in \mathds{R}$, \Eqref{eqn:dod_3d_uniaxial} remains a real-valued function. Hence, a distinction between the two cases for the diffusion coefficients is not required for the distribution of diffusivities.

In the uniaxial case the first three moments simplify to
\begin{equation}
    \label{eqn:m1_uniaxial_3d}
    \moment{1} = \frac{1}{3}(\multipleEigenvalueOfD{1} + 2 \multipleEigenvalueOfD{2})
    \text{,}
\end{equation}
\begin{equation}
    \label{eqn:m2_uniaxial_3d}
    \moment{2} = \frac{1}{9}\left(3 {\multipleEigenvalueOfD{1}}^2 + 4 {\multipleEigenvalueOfD{1}} {\multipleEigenvalueOfD{2}} + 8 {\multipleEigenvalueOfD{2}}^2\right)
\end{equation}
and
\begin{equation}
    \label{eqn:m3_uniaxial_3d}
    \moment{3} = \frac{1}{9}\left(5 {\multipleEigenvalueOfD{1}}^3 + 6 {\multipleEigenvalueOfD{1}}^2 {\multipleEigenvalueOfD{2}} + 8 {\multipleEigenvalueOfD{1}} {\multipleEigenvalueOfD{2}}^2 + 16 {\multipleEigenvalueOfD{2}}^3\right)
    \text{.}
\end{equation}
Thus, the eigenvalues of $\ten{D}$ are calculated by
\begin{equation}
  \label{eqn:multiple 1 eigenvalue in dependence on moments}
    \multipleEigenvalueOfD{1} = \moment{1} \mp \sqrt{3 \moment{2} - 5 \moment{1}^2}
\end{equation}
and
\begin{equation}
  \label{eqn:multiple 2 eigenvalue in dependence on moments}
    \multipleEigenvalueOfD{2} = \moment{1} \pm \frac{1}{2}\sqrt{3 \moment{2} - 5 \moment{1}^2}
\text{,}
\end{equation}
where the sign in the equations depends on the constraint of positive diffusion coefficients. None of the eigenvalues will become complex since with \EqsAndref{eqn:m1_uniaxial_3d}{eqn:m2_uniaxial_3d} the expression under the square root $3\moment{2} - 5\moment{1}^2 = \frac{4}{9} (\multipleEigenvalueOfD{1}-\multipleEigenvalueOfD{2})^2 > 0$ is always positive and, hence, $\moment{2} > \frac{5}{3} \moment{1}^2$. It should be noted that for $\frac{5}{3} \moment{1}^2 < \moment{2} < 2 \moment{1}^2$ both signs in \EqsAndref{eqn:multiple 1 eigenvalue in dependence on moments}{eqn:multiple 2 eigenvalue in dependence on moments} yield positive diffusion coefficients. In this case, the third moment has to be exploited in order to decide the correct pair of diffusion coefficients by comparing \Eqref{eqn:m3_uniaxial_3d} with the measured value. Hence, there exist distributions of diffusivities with identical moments $\moment{1}$ and $\moment{2}$, which result from different diffusion coefficients. In this case, the distinct $\moment{3}$ determines the corresponding diffusion coefficients of the system. In the limit $\moment{2} \to \frac{5}{3} \moment{1}^2$, $\multipleEigenvalueOfD{1}$ and $\multipleEigenvalueOfD{2}$ approach each other. In this particular case, the decision for the correct pair cannot be made accurately since both pairs yield approximately the same $\moment{3}$ from \Eqref{eqn:m3_uniaxial_3d}. However, this limit corresponds to the isotropic system and, hence, the single diffusion coefficient is directly given by the first moment of the distribution.

\begin{figure}[ht]
	\includegraphics{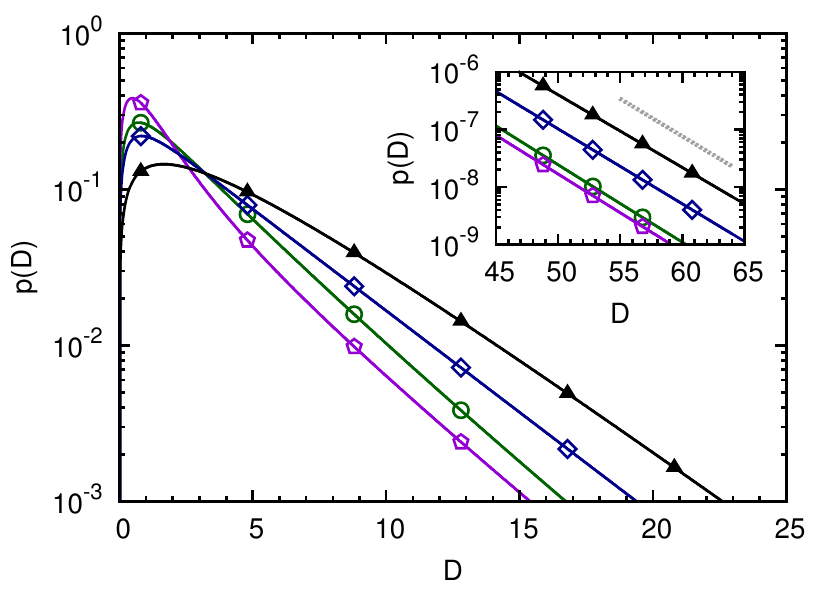}
\caption{\label{fig:dod of 3d uniaxial and biaxial anisotropic diffusion}Distribution of diffusivities (lines with open symbols) of different homogeneous anisotropic diffusion processes in three dimensions. A qualitative distinction between the oblate case (\myopendiamond{darkblue}; $\multipleEigenvalueOfD{1} = 1, \multipleEigenvalueOfD{2} = 5$), the prolate case (\myopenpentagon{darkviolet}; $\multipleEigenvalueOfD{1} = 5, \multipleEigenvalueOfD{2} = 1$), and a general anisotropic case (\myopencircle{darkgreen}; $D_1 = 5, D_2 = 3, D_3 = 1$) is possible since the decay after the maximum peak shows a concave curvature in the first case and a convex curvature in the latter cases. The inset shows that each anisotropic case obeys the same asymptotic decay given by the largest diffusion coefficient (dotted line as a guide to the eye). For comparison the isotropic case with the same asymptotic decay (\myfilledtriangle{black}; $D_c = 5$) is given, which always has a concave shape. Thus, it is qualitatively indistinguishable from the oblate case. However, a comparison of the first moment with the asymptotic decay offers a simple distinction between both cases.}
\end{figure}

\Figref{fig:dod of 3d uniaxial and biaxial anisotropic diffusion} depicts examples of such distributions for the general anisotropic, the prolate, and the oblate case. The differences can be identified qualitatively. In the general and in the prolate case, the decay after the maximum peak has a convex curvature in the logarithmic representation, whereas in the oblate case it decays in a purely concave manner. This qualitative change is obtained from $\frac{\dd^2}{\dd D^2} \log \puni{3}(D)$ and discussed in \Secref{sec:curvature}. In all cases, the exponential decay for large $D$ is determined by the largest diffusion coefficient as given by \Eqref{eqn:asymptotic_3d_aniso}. However, since the first decay after the peak is dominated by the smallest diffusion coefficient, the curve is shifted to the left for the prolate case in contrast to the oblate case when $D_2$ is changed from $D_3$ to $D_1$. As expected from the first moment, the general case lies in between. A better distinction between the different cases is achieved quantitatively by determining the moments and calculating the diffusion coefficients.

\begin{figure}[ht]
	\includegraphics{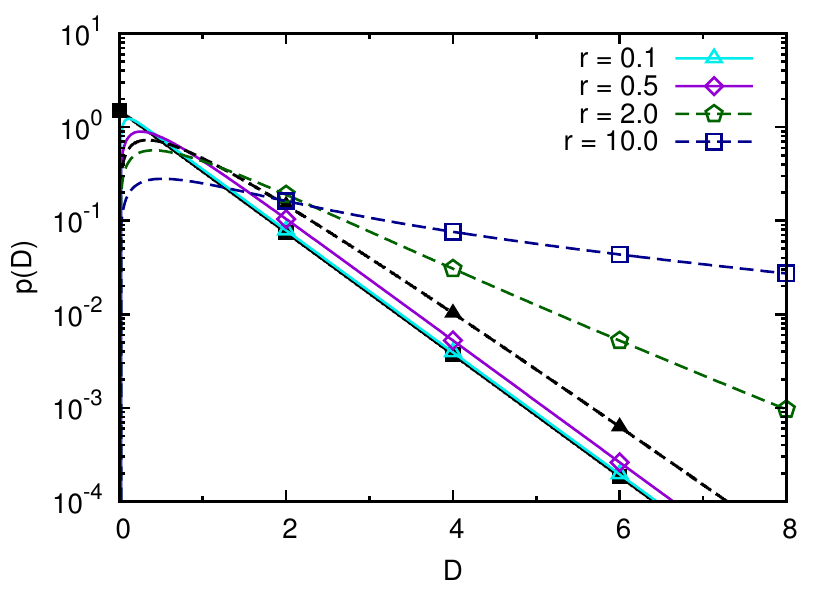}
\caption{\label{fig:dod of 3d uniaxial anisotropic diffusion for different ratios d1 d2}Distribution of diffusivities (lines with open symbols) for different ratios $r$ given by \Eqref{eqn:ratio of eigenvalues} and fixed $\multipleEigenvalueOfD{2}=1$. The crossover from oblate cases ($r<1$, solid lines) to prolate cases ($r>1$, dashed lines) shows a broadening of the peak for increasing ratios. Again, the behavior after the peak changes from concave to convex, respectively. For comparison, the distribution of diffusivities of the limiting isotropic cases are depicted for two-dimensional (\myfilledsquare{black}, $D_c = 2/3$) and three-dimensional processes (\myfilledtriangle{black}, $D_c = 1$). The distinction of prolate cases from the isotropic limits is simpler than for the oblate cases.}
\end{figure}

In \Figref{fig:dod of 3d uniaxial anisotropic diffusion for different ratios d1 d2} the distribution of diffusivities for different ratios
\begin{equation}
 \label{eqn:ratio of eigenvalues}
 r = \multipleEigenvalueOfD{1}/\multipleEigenvalueOfD{2}
\end{equation}
is shown, ranging from oblate cases ($r<1$) to prolate cases. It can be seen that in the limit $\multipleEigenvalueOfD{1} \to 0$ and, thus, $r \to 0$, the distribution converges to the two-dimensional isotropic case with $D_c = 2/3\multipleEigenvalueOfD{2}$. For $r \to 1$, the distribution converges to the three-dimensional isotropic case. In the prolate cases the distribution separates significantly from the three-dimensional isotropic case for increasing $r$. For further increasing ratios ($r \to \infty$) the distribution converges to the one-dimensional isotropic case with $D_c = 1/3 \multipleEigenvalueOfD{1}$. In contrast, the oblate cases converge rapidly to the two-dimensional isotropic case for decreasing $r$. A qualitative distinction may only be possible for small $D$, where the distribution still deviates from the mono-exponential behavior of the isotropic system. However, quantitatively the anisotropy is characterized by \Eqref{eqn:anisotropy measure from asymptotic decays}, which results in $\anisotropyMeasure=\tfrac{1-r}{2+r}$ and $\anisotropyMeasure=\tfrac{2(r-1)}{2+r}$ for oblate and prolate cases, respectively. Thus, in the oblate case the largest possible anisotropy emerges at small $r$, which yields $\anisotropyMeasure=1/2$ and clearly indicates the anisotropy. In the prolate case, the largest anisotropy will be obtained, if only one direction is preferred. Then, the anisotropy measure $\anisotropyMeasure=2$ is maximal, which corresponds to one-dimensional motion in a three-dimensional system.

\subsubsection{\label{sec:curvature}Curvature of the distribution of diffusivities}
As noticed in \Figref{fig:dod of 3d uniaxial and biaxial anisotropic diffusion}, the convex or concave curvature of the probability density in the logarithmic representation depends on the observed system and, thus, on the structure of the diffusion tensor. In the literature, such a concave curvature is known as log-concavity of functions which is a common property of probability distributions and has been studied extensively \cite{An1998,Bagnoli2005,Gupta2012}. However, in the case of log-convex functions there are much less properties known. In the following, we discuss the curvature of the distribution of diffusivities in the logarithmic representation, which can be exploited to determine characteristic properties of the observed processes.

For anisotropic processes the asymptotic curvature of the distribution of diffusivities in the logarithmic representation is obtained from the uniaxial case \Eqref{eqn:dod_3d_uniaxial} since \Eqref{eqn:dod_3d_aniso} does not provide a closed-form expression. For isotropic diffusion the curvature of the distribution of diffusivities is determined from \Eqref{eqn:dod 3d isotropic homogeneous}.

The asymptotic expansion of the second derivative for small $D$ yields
\begin{equation}
\label{eqn:curvature for small D}
 \frac{\dd^2}{\dd D^2} \log \paniso{3}(D) \overset{D \to 0}{\sim} -1/(2 D^2)
\text{,}
\end{equation}
which coincides with the curvature of three-dimensional isotropic systems. Analogously, we perform the asymptotic expansion of the second derivative for large $D$
\begin{align}
\label{eqn:curvature for large D}
& \frac{\dd^2}{\dd D^2} \log \paniso{3}(D) \nonumber\\
& \overset{D \to \infty}{\sim}
  \begin{cases}
	1/(2 D^2) 	& D_1 > D_2 = D_3, \\
	- \frac{a^{3/2}}{\sqrt{\pi D}} \exp(-a D) 	& D_1 = D_2 > D_3, \\
	-1/(2 D^2) 	& D_1 = D_2 = D_3
  \end{cases}
\end{align}
with positive $a = 3/2 (1/\multipleEigenvalueOfD{1} - 1/\multipleEigenvalueOfD{2})$. The different results depend on the multiplicity of the largest eigenvalue for prolate, oblate and isotropic cases, respectively. In the general anisotropic case with $D_1 \ne D_2 \ne D_3$, the system is dominated by the largest diffusion coefficient for large $D$. Hence, in this case the asymptotic curvature is identical to the prolate case ($D_1 > D_2 = D_3$) and can also be obtained from \Eqref{eqn:asymptotic_Nd_aniso}. As expected, a degeneracy of the smaller eigenvalues does not contribute to the asymptotic curvature. Hence, in all systems where the largest eigenvalue is not degenerated, for instance in anisotropic two-dimensional and also one-dimensional systems, we obtain the same behavior for large $D$, which is governed by the largest eigenvalue of the system.

The curvature of the distribution of diffusivities in the logarithmic representation for small $D$ is always concave as given by \Eqref{eqn:curvature for small D}. However, for large $D$ it depends on the observed system showing either a convex or a concave behavior as given in \Eqref{eqn:curvature for large D}. Hence, the sign of the curvature can change with $D$. In the prolate case the corresponding point of inflection is found to be approximately at $1.504 \multipleEigenvalueOfD{1}\multipleEigenvalueOfD{2}/(\multipleEigenvalueOfD{1}-\multipleEigenvalueOfD{2})$ by numerical evaluation of the root of $\frac{\dd^2}{\dd D^2} \log \paniso{3}(D)$. For anisotropic systems, only in the oblate case the curvature does not change its sign and the distribution is a log-concave function. If the anisotropy measure becomes zero and the distribution is a log-concave function, a three-dimensional isotropic diffusion process is observed. This qualitative difference in the curvature of distributions with the same asymptotic decay can clearly be identified in \Figref{fig:dod of 3d uniaxial and biaxial anisotropic diffusion}.

Furthermore, it is interesting to note in \Eqref{eqn:curvature for large D} that in the oblate case, where the largest eigenvalue exhibits a twofold degeneracy, the asymptotic behavior of the curvature still depends on the diffusion coefficients of the system. In all other cases the dependence on the diffusion coefficients vanishes.

As noted above, for two-dimensional anisotropic processes the asymptotic behavior for large $D$ in the logarithmic representation is identical to the prolate case in \Eqref{eqn:curvature for large D}. However, the asymptotic behavior for small $D$ is given by $1/8 (1/D_1 - 1/D_2)^2$ and clearly differs from that of the three-dimensional process. Since the sign of the curvature does not change with $D$ the curvature is always convex in two-dimensional anisotropic systems. However, for two-dimensional isotropic systems the distribution of diffusivities in the logarithmic representation is just a straight line for all $D$.

\subsubsection{\label{sec:reconstruction3d}Reconstruction of \texorpdfstring{$\ten{D}$}{D}}
As discussed for two-dimensional processes, the diffusion tensor will be easily obtained by measuring the averaged tensorial diffusivities according to \Eqref{eqn:tensor estimation} if the complete three-dimensional trajectory of the homogeneous anisotropic process is available. For the sample trajectory used in \Figref{fig:dod of 3d_biaxial anisotropic diffusion} the measured values $\measured{D}_{ij}$ yield the diffusion tensor
\begin{equation}
    \measured{\ten{D}} = \begin{pmatrix}
                            3.994 & -0.862 & -0.492 \\
                            -0.862 &  3.233 &  1.296 \\
                            -0.492 &  1.296 &  1.758
                         \end{pmatrix}
\text{,}
\end{equation}
which agrees reasonably with the input parameters of the simulation \Eqref{eqn:simulation_tensor_3d}. Further, the eigenvalues from this measured tensor $\measured{D}_{1} = 4.983$, $\measured{D}_{2} = 2.998$ and $\measured{D}_{3} = 1.004$ show a good agreement with the exact eigenvalues of the input tensor $D_1 = 5$, $D_2 = 3$ and $D_3 = 1$.

However, if only a projection of the complete trajectory is available, e.g.\ from SPT, only the properties of the respective submatrix of $\ten{D}$ can be measured. For instance, if the two-dimensional projection onto the x-y-plane of the sample trajectory is available, the first two moments of the distribution of diffusivities are determined to be $\measured{M}_{1,z} = 3.613$ and $\measured{M}_{2,z} = 26.95$. Using \Eqref{eqn:eigenvalues2d}, the eigenvalues of the principal submatrix $\ten{D}_{z}^2$ are computed to be $\measured{D}_{1,z}^2 = 4.531$ and $\measured{D}_{2,z}^2 = 2.695$. Hence, the eigenvalue inequalities of \Eqref{eqn:cauchy interlacing inequalities} provide the estimate
\begin{equation}
    D_1 \geq \measured{D}_{1,z}^2 = 4.531 \geq D_2 \geq \measured{D}_{2,z}^2 = 2.695 \geq D_3 \geq 0 \text{.}
\end{equation}
of the diffusion coefficients. As explained in \Secref{sec:Projection}, any further observed projection improves the estimates of the eigenvalues of $\ten{D}$. An additional projection onto the x-z-plane, for instance, yields the moments $\measured{M}_{1,y} = 2.876$ and $\measured{M}_{2,y} = 18.00$ resulting in the eigenvalues $\measured{D}_{1,y}^2 = 4.083$ and $\measured{D}_{2,y}^2 = 1.668$. Since with two orthogonal two-dimensional projections of the three-dimensional process all diagonal elements of $\ten{D}$ are available, an upper bound for the largest eigenvalue is found to be $D_1 \leq \tr \ten{D} \leq \measured{D}_{1,z}^2 + \measured{D}_{2,z}^2 + \measured{D}_{1,y}^2 + \measured{D}_{2,y}^2 = 12.977$. Hence, the eigenvalue inequalities yield
\begin{align}
    12.977 & \geq D_1 \geq \max(\measured{D}_{1,z}^2, \measured{D}_{1,y}^2) = 4.531 \nonumber\\
    \min(\measured{D}_{1,z}^2, \measured{D}_{1,y}^2) = 4.083 & \geq D_2 \geq  \max(\measured{D}_{2,z}^2, \measured{D}_{2,y}^2) = 2.695 \nonumber\\
    \min(\measured{D}_{2,z}^2, \measured{D}_{2,y}^2) = 1.668 & \geq D_3 \geq 0 \text{.}
\end{align}
If additionally the projection onto the y-z-plane is available the eigenvalues of $\ten{D}$ are estimated more precisely similar to the previous steps. To improve the upper bound of $D_1$, the trace of $\ten{D}$ is calculated from all these eigenvalues by $\tr\ten{D} = \onehalf(\measured{D}_{1,x}^2 + \measured{D}_{2,x}^2 + \measured{D}_{1,y}^2 + \measured{D}_{2,y}^2 + \measured{D}_{1,z}^2 + \measured{D}_{2,z}^2)$, where the prefactor arises from the overlapping diagonal elements of the submatrices.

In the case of availability of all orthogonal two-dimensional projections of the process the tensorial diffusivities offer an advanced approach to determine the diffusion tensor. Since their first moments yield the entries of the principal submatrices $\ten{D}_{x}^2$, $\ten{D}_{y}^2$ and $\ten{D}_{z}^2$ the underlying diffusion tensor $\ten{D}$ is completely defined.

To summarize, the experimental setup influences the available data and affects how many parameters of the underlying process can be restored. A single two-dimensional projection may already hint at the anisotropy of the process. However, it is not sufficient to give an upper bound for the largest eigenvalue. An additional orthogonal two-dimensional projection or even a one-dimensional projection in the missing direction determines this upper bound and narrows the ranges of the eigenvalues. For a reconstruction of the complete tensor either the complete trajectory or three orthogonal two-dimensional projections of the process are necessary.

%***The Chapman-Kolmogorov equation is valid, thus, even the projected process of anisotropic diffusion is Markovian.

%%%%%%%%%%%%%%%%%%%%%%%%%%%

\section{\label{sec:Conclusions}Conclusions}

To investigate $N$-dimensional homogeneous anisotropic Brownian motion we applied the distribution of diffusivities as e.g.\ obtained from single-particle tracking data. We introduced an anisotropy measure depending on the asymptotic decay of the distribution and the mean of the diffusivities, which both are easily determined from experimental data. In general, if this anisotropy measure is larger than zero, the distribution deviates from the $\chi^2$-distribution, which we obtain for homogeneous isotropic diffusion. Thus, the observed process involves more than one diffusion coefficient attributed to an inhomogeneity or an anisotropy of the system. For homogeneous processes we concluded that those systems have to be anisotropic. Furthermore, from the general expression of the distribution of diffusivities we derived relations between its moments or cumulants and the eigenvalues of the diffusion tensor $\ten{D}$. Since, due to experimental restrictions, often only projections of the trajectories are observed we further discussed the consequences and provided an estimate for the bounds of the involved diffusion coefficients.

After our general considerations, we applied the results to specific systems with high relevance to experiments. In particular, we investigated two-dimensional and three-dimensional systems as well as uniaxial molecules in three dimensions. In a two-dimensional homogeneous anisotropic system, the distribution of diffusivities comprises a modified Bessel function and allows a qualitative distinction from the mono-exponential decay observed in isotropic systems. Moreover, the first two moments of the distribution are sufficient to calculate the diffusion coefficients corresponding to the principal axes. Even the orientation of the principal axes and, thus, the complete diffusion tensor $\ten{D}$ can be determined by using tensorial diffusivities or three one-dimensional projections of the trajectory. For three-dimensional processes the general expression of the distribution of diffusivities is more elaborated and one integration has to be evaluated numerically. However, we expressed the first three moments in terms of the diffusion coefficients belonging to the principal axes. Conversely, these expressions offer a method to calculate the diffusion coefficients from the moments measured in experiments, where other analysis fails. It is further shown that the isotropic and anisotropic systems differ in the logarithmic representation of the distribution of diffusivities, i.e., the asymptotic decay rate is proportional to the inverse slope of the msd and to the inverse of the largest diffusion coefficient, respectively. Thus, the distribution of diffusivities for anisotropic diffusion asymptotically decays slower than for isotropic diffusion with the same mean diffusion coefficient. The deviation between the asymptotic decay and the first moment provides a suitable measure for the anisotropy of the process. For uniaxial molecules diffusing in three dimensions the third integration was accomplished and the resulting distribution of diffusivities involves an error function. In this case, the diffusion coefficients along the direction of the principal axes depend on the first two moments of the distribution. For different ratios of the diffusion coefficients we distinguish between oblate and prolate cases, which show a concave and a convex curvature in the logarithmic representation, respectively. Finally, we offer a guide to quantify the eigenvalues of $\ten{D}$ from projected observations and to reconstruct the diffusion tensor in three dimensions from the moments of the tensorial diffusivities. The reconstruction from projected observations is possible although any directional information is discarded when determining the distribution of diffusivities.

In summary, the distribution of diffusivities provides an advanced analysis of anisotropic diffusion processes. The distribution is easily obtained from measured trajectories or from ensemble measurements such as NMR and allows for a characterization of the processes. For time-homogeneous diffusion processes, this distribution is stationary, which allows us to compare experiments conducted on different time scales. The first moment of the distribution corresponds to the mean of the diffusivities and coincides with the slope of the mean squared displacement. From the discrepancy between the asymptotic decay of the distribution and the mean of the diffusivities it is easy to identify systems which are not sufficiently characterized by a single diffusion coefficient. Hence, we encourage experimentalists to determine these simple quantities in order to detect a discrepancy and to verify their assumptions about homogeneous isotropic processes. Furthermore, if the system is homogeneous and anisotropic, the diffusion coefficients can be reconstructed from the moments of the distribution. Beyond that, the concept of diffusivities as scaled displacements is extended to tensorial diffusivities, which allow the reconstruction of the diffusion tensor from their first moments. Hence, the distribution of diffusivities complements well-established methods, such as investigating mean squared displacements, for the analysis of diffusion data.

In future publications, we will address the distinction between anisotropic and heterogeneous diffusion processes, which also involve more than one diffusion coefficient. Furthermore, since the eigenvalues of the tensor $\ten{D}$ are invariant to orthogonal transformations, we will apply our distribution of diffusivities to systems where the diffusion tensor changes its orientation in space and time, such as diffusion of ellipsoidal particles in isotropic media and diffusion in liquid crystalline systems with an inhomogeneous director field.

\begin{acknowledgments}
We thank Sven Schubert for stimulating discussions and valuable suggestions. We gratefully acknowledge financial support from the Deutsche Forschungsgemeinschaft (DFG) for funding of the research unit FOR 877 ``From Local Constraints to Macroscopic Transport''.
\end{acknowledgments}

%\appendix
%\section{\label{sec:Appendix}Appendix}

%\section{\label{sec:Appendix Test}Testappendix}

%\section{ToDos}

\bibliography{references}

%merlin.mbs apsrev4-1.bst 2010-07-25 4.21a (PWD, AO, DPC) hacked
%Control: key (0)
%Control: author (8) initials jnrlst
%Control: editor formatted (1) identically to author
%Control: production of article title (-1) disabled
%Control: page (0) single
%Control: year (1) truncated
%Control: production of eprint (0) enabled
\begin{thebibliography}{35}%
\makeatletter
\providecommand \@ifxundefined [1]{%
 \@ifx{#1\undefined}
}%
\providecommand \@ifnum [1]{%
 \ifnum #1\expandafter \@firstoftwo
 \else \expandafter \@secondoftwo
 \fi
}%
\providecommand \@ifx [1]{%
 \ifx #1\expandafter \@firstoftwo
 \else \expandafter \@secondoftwo
 \fi
}%
\providecommand \natexlab [1]{#1}%
\providecommand \enquote  [1]{``#1''}%
\providecommand \bibnamefont  [1]{#1}%
\providecommand \bibfnamefont [1]{#1}%
\providecommand \citenamefont [1]{#1}%
\providecommand \href@noop [0]{\@secondoftwo}%
\providecommand \href [0]{\begingroup \@sanitize@url \@href}%
\providecommand \@href[1]{\@@startlink{#1}\@@href}%
\providecommand \@@href[1]{\endgroup#1\@@endlink}%
\providecommand \@sanitize@url [0]{\catcode `\\12\catcode `\$12\catcode
  `\&12\catcode `\#12\catcode `\^12\catcode `\_12\catcode `\%12\relax}%
\providecommand \@@startlink[1]{}%
\providecommand \@@endlink[0]{}%
\providecommand \url  [0]{\begingroup\@sanitize@url \@url }%
\providecommand \@url [1]{\endgroup\@href {#1}{\urlprefix }}%
\providecommand \urlprefix  [0]{URL }%
\providecommand \Eprint [0]{\href }%
\providecommand \doibase [0]{http://dx.doi.org/}%
\providecommand \selectlanguage [0]{\@gobble}%
\providecommand \bibinfo  [0]{\@secondoftwo}%
\providecommand \bibfield  [0]{\@secondoftwo}%
\providecommand \translation [1]{[#1]}%
\providecommand \BibitemOpen [0]{}%
\providecommand \bibitemStop [0]{}%
\providecommand \bibitemNoStop [0]{.\EOS\space}%
\providecommand \EOS [0]{\spacefactor3000\relax}%
\providecommand \BibitemShut  [1]{\csname bibitem#1\endcsname}%
\let\auto@bib@innerbib\@empty
%</preamble>
\bibitem [{\citenamefont {van Kampen}(1988)}]{vanKampen1988673}%
  \BibitemOpen
  \bibfield  {author} {\bibinfo {author} {\bibfnamefont {N.~G.}\ \bibnamefont
  {van Kampen}},\ }\href {\doibase 10.1016/0022-3697(88)90199-0} {\bibfield
  {journal} {\bibinfo  {journal} {J. Phys. Chem. Solids}\ }\textbf {\bibinfo
  {volume} {49}},\ \bibinfo {pages} {673} (\bibinfo {year} {1988})}\BibitemShut
  {NoStop}%
\bibitem [{\citenamefont {Christensen}\ and\ \citenamefont
  {Pedersen}(2003)}]{Christensen20035171}%
  \BibitemOpen
  \bibfield  {author} {\bibinfo {author} {\bibfnamefont {M.}~\bibnamefont
  {Christensen}}\ and\ \bibinfo {author} {\bibfnamefont {J.~B.}\ \bibnamefont
  {Pedersen}},\ }\href {\doibase 10.1063/1.1597476} {\bibfield  {journal}
  {\bibinfo  {journal} {J. Chem. Phys.}\ }\textbf {\bibinfo {volume} {119}},\
  \bibinfo {pages} {5171} (\bibinfo {year} {2003})}\BibitemShut {NoStop}%
\bibitem [{\citenamefont {Bringuier}(2011)}]{Bringuier2011}%
  \BibitemOpen
  \bibfield  {author} {\bibinfo {author} {\bibfnamefont {E.}~\bibnamefont
  {Bringuier}},\ }\href {\doibase 10.1088/0143-0807/32/4/012} {\bibfield
  {journal} {\bibinfo  {journal} {Eur. J. Phys.}\ }\textbf {\bibinfo {volume}
  {32}},\ \bibinfo {pages} {975} (\bibinfo {year} {2011})}\BibitemShut
  {NoStop}%
\bibitem [{\citenamefont {Schulz}\ \emph {et~al.}(2010)\citenamefont {Schulz},
  \citenamefont {Täuber}, \citenamefont {Friedriszik}, \citenamefont {Graaf},
  \citenamefont {Schuster},\ and\ \citenamefont {von
  Borczyskowski}}]{schulz2010}%
  \BibitemOpen
  \bibfield  {author} {\bibinfo {author} {\bibfnamefont {B.}~\bibnamefont
  {Schulz}}, \bibinfo {author} {\bibfnamefont {D.}~\bibnamefont {Täuber}},
  \bibinfo {author} {\bibfnamefont {F.}~\bibnamefont {Friedriszik}}, \bibinfo
  {author} {\bibfnamefont {H.}~\bibnamefont {Graaf}}, \bibinfo {author}
  {\bibfnamefont {J.}~\bibnamefont {Schuster}}, \ and\ \bibinfo {author}
  {\bibfnamefont {C.}~\bibnamefont {von Borczyskowski}},\ }\href {\doibase
  10.1039/C004042H} {\bibfield  {journal} {\bibinfo  {journal} {Phys. Chem.
  Chem. Phys.}\ }\textbf {\bibinfo {volume} {12}},\ \bibinfo {pages} {11555}
  (\bibinfo {year} {2010})}\BibitemShut {NoStop}%
\bibitem [{\citenamefont {Schulz}\ \emph {et~al.}(2011)\citenamefont {Schulz},
  \citenamefont {Täuber}, \citenamefont {Schuster}, \citenamefont
  {Baumgärtel},\ and\ \citenamefont {von Borczyskowski}}]{schulz2011}%
  \BibitemOpen
  \bibfield  {author} {\bibinfo {author} {\bibfnamefont {B.}~\bibnamefont
  {Schulz}}, \bibinfo {author} {\bibfnamefont {D.}~\bibnamefont {Täuber}},
  \bibinfo {author} {\bibfnamefont {J.}~\bibnamefont {Schuster}}, \bibinfo
  {author} {\bibfnamefont {T.}~\bibnamefont {Baumgärtel}}, \ and\ \bibinfo
  {author} {\bibfnamefont {C.}~\bibnamefont {von Borczyskowski}},\ }\href
  {\doibase 10.1039/c1sm05434a} {\bibfield  {journal} {\bibinfo  {journal}
  {Soft Matter}\ }\textbf {\bibinfo {volume} {7}},\ \bibinfo {pages} {7431}
  (\bibinfo {year} {2011})}\BibitemShut {NoStop}%
\bibitem [{\citenamefont {Bechhoefer}\ \emph {et~al.}(1997)\citenamefont
  {Bechhoefer}, \citenamefont {Géminard}, \citenamefont {Bocquet},\ and\
  \citenamefont {Oswald}}]{PhysRevLett.79.4922}%
  \BibitemOpen
  \bibfield  {author} {\bibinfo {author} {\bibfnamefont {J.}~\bibnamefont
  {Bechhoefer}}, \bibinfo {author} {\bibfnamefont {J.-C.}\ \bibnamefont
  {Géminard}}, \bibinfo {author} {\bibfnamefont {L.}~\bibnamefont {Bocquet}},
  \ and\ \bibinfo {author} {\bibfnamefont {P.}~\bibnamefont {Oswald}},\ }\href
  {\doibase 10.1103/PhysRevLett.79.4922} {\bibfield  {journal} {\bibinfo
  {journal} {Phys. Rev. Lett.}\ }\textbf {\bibinfo {volume} {79}},\ \bibinfo
  {pages} {4922} (\bibinfo {year} {1997})}\BibitemShut {NoStop}%
\bibitem [{\citenamefont {Wax}(1954)}]{wax1954}%
  \BibitemOpen
  \bibinfo {editor} {\bibfnamefont {N.}~\bibnamefont {Wax}},\ ed.,\ \href@noop
  {} {\emph {\bibinfo {title} {Selected Papers on Noise and Stochastic
  Processes}}}\ (\bibinfo  {publisher} {Dover Publications},\ \bibinfo
  {address} {New York},\ \bibinfo {year} {1954})\BibitemShut {NoStop}%
\bibitem [{\citenamefont {Saxton}\ and\ \citenamefont
  {Jacobson}(1997)}]{SaxtonAnnuRef1997}%
  \BibitemOpen
  \bibfield  {author} {\bibinfo {author} {\bibfnamefont {M.~J.}\ \bibnamefont
  {Saxton}}\ and\ \bibinfo {author} {\bibfnamefont {K.}~\bibnamefont
  {Jacobson}},\ }\href {\doibase 10.1146/annurev.biophys.26.1.373} {\bibfield
  {journal} {\bibinfo  {journal} {Annu. Rev. Biophys. Biomol. Struct.}\
  }\textbf {\bibinfo {volume} {26}},\ \bibinfo {pages} {373} (\bibinfo {year}
  {1997})}\BibitemShut {NoStop}%
\bibitem [{\citenamefont {Dahan}\ \emph {et~al.}(2003)\citenamefont {Dahan},
  \citenamefont {Lévi}, \citenamefont {Luccardini}, \citenamefont {Rostaing},
  \citenamefont {Riveau},\ and\ \citenamefont {Triller}}]{Dahan17102003}%
  \BibitemOpen
  \bibfield  {author} {\bibinfo {author} {\bibfnamefont {M.}~\bibnamefont
  {Dahan}}, \bibinfo {author} {\bibfnamefont {S.}~\bibnamefont {Lévi}},
  \bibinfo {author} {\bibfnamefont {C.}~\bibnamefont {Luccardini}}, \bibinfo
  {author} {\bibfnamefont {P.}~\bibnamefont {Rostaing}}, \bibinfo {author}
  {\bibfnamefont {B.}~\bibnamefont {Riveau}}, \ and\ \bibinfo {author}
  {\bibfnamefont {A.}~\bibnamefont {Triller}},\ }\href {\doibase
  10.1126/science.1088525} {\bibfield  {journal} {\bibinfo  {journal}
  {Science}\ }\textbf {\bibinfo {volume} {302}},\ \bibinfo {pages} {442}
  (\bibinfo {year} {2003})}\BibitemShut {NoStop}%
\bibitem [{\citenamefont {Chen}\ \emph {et~al.}(2003)\citenamefont {Chen},
  \citenamefont {Weeks}, \citenamefont {Crocker}, \citenamefont {Islam},
  \citenamefont {Verma}, \citenamefont {Gruber}, \citenamefont {Levine},
  \citenamefont {Lubensky},\ and\ \citenamefont
  {Yodh}}]{PhysRevLett.90.108301}%
  \BibitemOpen
  \bibfield  {author} {\bibinfo {author} {\bibfnamefont {D.~T.}\ \bibnamefont
  {Chen}}, \bibinfo {author} {\bibfnamefont {E.~R.}\ \bibnamefont {Weeks}},
  \bibinfo {author} {\bibfnamefont {J.~C.}\ \bibnamefont {Crocker}}, \bibinfo
  {author} {\bibfnamefont {M.~F.}\ \bibnamefont {Islam}}, \bibinfo {author}
  {\bibfnamefont {R.}~\bibnamefont {Verma}}, \bibinfo {author} {\bibfnamefont
  {J.}~\bibnamefont {Gruber}}, \bibinfo {author} {\bibfnamefont {A.~J.}\
  \bibnamefont {Levine}}, \bibinfo {author} {\bibfnamefont {T.~C.}\
  \bibnamefont {Lubensky}}, \ and\ \bibinfo {author} {\bibfnamefont {A.~G.}\
  \bibnamefont {Yodh}},\ }\href {\doibase 10.1103/PhysRevLett.90.108301}
  {\bibfield  {journal} {\bibinfo  {journal} {Phys. Rev. Lett.}\ }\textbf
  {\bibinfo {volume} {90}},\ \bibinfo {pages} {108301} (\bibinfo {year}
  {2003})}\BibitemShut {NoStop}%
\bibitem [{\citenamefont {Mason}\ \emph {et~al.}(1997)\citenamefont {Mason},
  \citenamefont {Ganesan}, \citenamefont {van Zanten}, \citenamefont {Wirtz},\
  and\ \citenamefont {Kuo}}]{PhysRevLett.79.3282}%
  \BibitemOpen
  \bibfield  {author} {\bibinfo {author} {\bibfnamefont {T.~G.}\ \bibnamefont
  {Mason}}, \bibinfo {author} {\bibfnamefont {K.}~\bibnamefont {Ganesan}},
  \bibinfo {author} {\bibfnamefont {J.~H.}\ \bibnamefont {van Zanten}},
  \bibinfo {author} {\bibfnamefont {D.}~\bibnamefont {Wirtz}}, \ and\ \bibinfo
  {author} {\bibfnamefont {S.~C.}\ \bibnamefont {Kuo}},\ }\href {\doibase
  10.1103/PhysRevLett.79.3282} {\bibfield  {journal} {\bibinfo  {journal}
  {Phys. Rev. Lett.}\ }\textbf {\bibinfo {volume} {79}},\ \bibinfo {pages}
  {3282} (\bibinfo {year} {1997})}\BibitemShut {NoStop}%
\bibitem [{\citenamefont {McHale}\ \emph {et~al.}(2007)\citenamefont {McHale},
  \citenamefont {Berglund},\ and\ \citenamefont {Mabuchi}}]{NanoLett.7.11}%
  \BibitemOpen
  \bibfield  {author} {\bibinfo {author} {\bibfnamefont {K.}~\bibnamefont
  {McHale}}, \bibinfo {author} {\bibfnamefont {A.~J.}\ \bibnamefont
  {Berglund}}, \ and\ \bibinfo {author} {\bibfnamefont {H.}~\bibnamefont
  {Mabuchi}},\ }\href {\doibase 10.1021/nl0723376} {\bibfield  {journal}
  {\bibinfo  {journal} {Nano Lett.}\ }\textbf {\bibinfo {volume} {7}},\
  \bibinfo {pages} {3535} (\bibinfo {year} {2007})}\BibitemShut {NoStop}%
\bibitem [{\citenamefont {Wells}\ \emph {et~al.}(2008)\citenamefont {Wells},
  \citenamefont {Lessard},\ and\ \citenamefont {Werner}}]{An.Chem.80.24}%
  \BibitemOpen
  \bibfield  {author} {\bibinfo {author} {\bibfnamefont {N.~P.}\ \bibnamefont
  {Wells}}, \bibinfo {author} {\bibfnamefont {G.~A.}\ \bibnamefont {Lessard}},
  \ and\ \bibinfo {author} {\bibfnamefont {J.~H.}\ \bibnamefont {Werner}},\
  }\href {\doibase 10.1021/ac8021899} {\bibfield  {journal} {\bibinfo
  {journal} {Anal. Chem.}\ }\textbf {\bibinfo {volume} {80}},\ \bibinfo {pages}
  {9830} (\bibinfo {year} {2008})}\BibitemShut {NoStop}%
\bibitem [{\citenamefont {Spille}\ \emph {et~al.}(2012)\citenamefont {Spille},
  \citenamefont {Kaminski}, \citenamefont {Königshoven},\ and\ \citenamefont
  {Kubitscheck}}]{spille2012}%
  \BibitemOpen
  \bibfield  {author} {\bibinfo {author} {\bibfnamefont {J.-H.}\ \bibnamefont
  {Spille}}, \bibinfo {author} {\bibfnamefont {T.}~\bibnamefont {Kaminski}},
  \bibinfo {author} {\bibfnamefont {H.-P.}\ \bibnamefont {Königshoven}}, \
  and\ \bibinfo {author} {\bibfnamefont {U.}~\bibnamefont {Kubitscheck}},\
  }\href {\doibase 10.1364/OE.20.019697} {\bibfield  {journal} {\bibinfo
  {journal} {Opt. Express}\ }\textbf {\bibinfo {volume} {20}},\ \bibinfo
  {pages} {19697} (\bibinfo {year} {2012})}\BibitemShut {NoStop}%
\bibitem [{\citenamefont {Ribrault}\ \emph {et~al.}(2007)\citenamefont
  {Ribrault}, \citenamefont {Triller},\ and\ \citenamefont
  {Sekimoto}}]{PhysRevE.75.021112}%
  \BibitemOpen
  \bibfield  {author} {\bibinfo {author} {\bibfnamefont {C.}~\bibnamefont
  {Ribrault}}, \bibinfo {author} {\bibfnamefont {A.}~\bibnamefont {Triller}}, \
  and\ \bibinfo {author} {\bibfnamefont {K.}~\bibnamefont {Sekimoto}},\ }\href
  {\doibase 10.1103/PhysRevE.75.021112} {\bibfield  {journal} {\bibinfo
  {journal} {Phys. Rev. E}\ }\textbf {\bibinfo {volume} {75}},\ \bibinfo
  {pages} {021112} (\bibinfo {year} {2007})}\BibitemShut {NoStop}%
\bibitem [{\citenamefont {Kärger}\ \emph {et~al.}(2012)\citenamefont
  {Kärger}, \citenamefont {Ruthven},\ and\ \citenamefont
  {Theodorou}}]{kaerger2012}%
  \BibitemOpen
  \bibfield  {author} {\bibinfo {author} {\bibfnamefont {J.}~\bibnamefont
  {Kärger}}, \bibinfo {author} {\bibfnamefont {D.~M.}\ \bibnamefont
  {Ruthven}}, \ and\ \bibinfo {author} {\bibfnamefont {D.~N.}\ \bibnamefont
  {Theodorou}},\ }\href@noop {} {\emph {\bibinfo {title} {Diffusion in
  Nanoporous Materials}}}\ (\bibinfo  {publisher} {Wiley-VCH},\ \bibinfo
  {address} {Weinheim},\ \bibinfo {year} {2012})\BibitemShut {NoStop}%
\bibitem [{\citenamefont {Hanasaki}\ and\ \citenamefont
  {Isono}(2012)}]{hanasaki2012}%
  \BibitemOpen
  \bibfield  {author} {\bibinfo {author} {\bibfnamefont {I.}~\bibnamefont
  {Hanasaki}}\ and\ \bibinfo {author} {\bibfnamefont {Y.}~\bibnamefont
  {Isono}},\ }\href {\doibase 10.1103/PhysRevE.85.051134} {\bibfield  {journal}
  {\bibinfo  {journal} {Physical Review E}\ }\textbf {\bibinfo {volume} {85}},\
  \bibinfo {pages} {051134} (\bibinfo {year} {2012})}\BibitemShut {NoStop}%
\bibitem [{\citenamefont {Bauer}\ \emph {et~al.}(2011)\citenamefont {Bauer},
  \citenamefont {Valiullin}, \citenamefont {Radons},\ and\ \citenamefont
  {Kärger}}]{bauer2011}%
  \BibitemOpen
  \bibfield  {author} {\bibinfo {author} {\bibfnamefont {M.}~\bibnamefont
  {Bauer}}, \bibinfo {author} {\bibfnamefont {R.}~\bibnamefont {Valiullin}},
  \bibinfo {author} {\bibfnamefont {G.}~\bibnamefont {Radons}}, \ and\ \bibinfo
  {author} {\bibfnamefont {J.}~\bibnamefont {Kärger}},\ }\href {\doibase
  10.1063/1.3647875} {\bibfield  {journal} {\bibinfo  {journal} {J. Chem.
  Phys.}\ }\textbf {\bibinfo {volume} {135}},\ \bibinfo {pages} {144118}
  (\bibinfo {year} {2011})}\BibitemShut {NoStop}%
\bibitem [{\citenamefont {Hellriegel}\ \emph {et~al.}(2005)\citenamefont
  {Hellriegel}, \citenamefont {Kirstein},\ and\ \citenamefont
  {Bräuchle}}]{hellriegel2005}%
  \BibitemOpen
  \bibfield  {author} {\bibinfo {author} {\bibfnamefont {C.}~\bibnamefont
  {Hellriegel}}, \bibinfo {author} {\bibfnamefont {J.}~\bibnamefont
  {Kirstein}}, \ and\ \bibinfo {author} {\bibfnamefont {C.}~\bibnamefont
  {Bräuchle}},\ }\href {\doibase 10.1088/1367-2630/7/1/023} {\bibfield
  {journal} {\bibinfo  {journal} {New J. Phys.}\ }\textbf {\bibinfo {volume}
  {7}},\ \bibinfo {pages} {23} (\bibinfo {year} {2005})}\BibitemShut {NoStop}%
\bibitem [{\citenamefont {H\"ofling}\ and\ \citenamefont
  {Franosch}(2013)}]{hoefling2013}%
  \BibitemOpen
  \bibfield  {author} {\bibinfo {author} {\bibfnamefont {F.}~\bibnamefont
  {H\"ofling}}\ and\ \bibinfo {author} {\bibfnamefont {T.}~\bibnamefont
  {Franosch}},\ }\href {\doibase 10.1088/0034-4885/76/4/046602} {\bibfield
  {journal} {\bibinfo  {journal} {Reports on Progress in Physics}\ }\textbf
  {\bibinfo {volume} {76}},\ \bibinfo {pages} {046602:1} (\bibinfo {year}
  {2013})}\BibitemShut {NoStop}%
\bibitem [{\citenamefont {Albers}\ and\ \citenamefont
  {Radons}(2013)}]{albers2013}%
  \BibitemOpen
  \bibfield  {author} {\bibinfo {author} {\bibfnamefont {T.}~\bibnamefont
  {Albers}}\ and\ \bibinfo {author} {\bibfnamefont {G.}~\bibnamefont
  {Radons}},\ }\href {\doibase 10.1209/0295-5075/102/40006} {\bibfield
  {journal} {\bibinfo  {journal} {EPL}\ }\textbf {\bibinfo {volume} {102}},\
  \bibinfo {pages} {40006} (\bibinfo {year} {2013})}\BibitemShut {NoStop}%
\bibitem [{\citenamefont {de~Gennes}\ and\ \citenamefont
  {Prost}(1995)}]{deGennes_B95}%
  \BibitemOpen
  \bibfield  {author} {\bibinfo {author} {\bibfnamefont {P.-G.}\ \bibnamefont
  {de~Gennes}}\ and\ \bibinfo {author} {\bibfnamefont {J.}~\bibnamefont
  {Prost}},\ }\href@noop {} {\emph {\bibinfo {title} {The Physics of Liquid
  Crystals}}},\ \bibinfo {edition} {2nd}\ ed.,\ \bibinfo {series}
  {International Series of Monographs on Physics}, Vol.~\bibinfo {volume} {83}\
  (\bibinfo  {publisher} {Clarendon Press},\ \bibinfo {address} {Oxford},\
  \bibinfo {year} {1995})\BibitemShut {NoStop}%
\bibitem [{\citenamefont {Risken}(1989)}]{Risken1989}%
  \BibitemOpen
  \bibfield  {author} {\bibinfo {author} {\bibfnamefont {H.}~\bibnamefont
  {Risken}},\ }\href@noop {} {\emph {\bibinfo {title} {The Fokker-Planck
  Equation: Methods of Solution and Applications}}},\ \bibinfo {edition} {2nd}\
  ed.,\ \bibinfo {series} {Springer Series in Synergetics}, Vol.~\bibinfo
  {volume} {18}\ (\bibinfo  {publisher} {Springer},\ \bibinfo {address}
  {Berlin},\ \bibinfo {year} {1989})\BibitemShut {NoStop}%
\bibitem [{\citenamefont {Price}(2009)}]{price2009}%
  \BibitemOpen
  \bibfield  {author} {\bibinfo {author} {\bibfnamefont {W.~S.}\ \bibnamefont
  {Price}},\ }\href {\doibase 10.1017/CBO9780511770487} {\emph {\bibinfo
  {title} {NMR Studies of Translational Motion}}},\ Cambridge Molecular
  Science\ (\bibinfo  {publisher} {Cambridge University Press},\ \bibinfo
  {address} {Cambridge, New York},\ \bibinfo {year} {2009})\BibitemShut
  {NoStop}%
\bibitem [{\citenamefont {Saxton}(1997)}]{saxton1997}%
  \BibitemOpen
  \bibfield  {author} {\bibinfo {author} {\bibfnamefont {M.~J.}\ \bibnamefont
  {Saxton}},\ }\href {\doibase 10.1016/S0006-3495(97)78820-9} {\bibfield
  {journal} {\bibinfo  {journal} {Biophys. J.}\ }\textbf {\bibinfo {volume}
  {72}},\ \bibinfo {pages} {1744} (\bibinfo {year} {1997})}\BibitemShut
  {NoStop}%
\bibitem [{\citenamefont {Smith}(1995)}]{Smith1995}%
  \BibitemOpen
  \bibfield  {author} {\bibinfo {author} {\bibfnamefont {P.~J.}\ \bibnamefont
  {Smith}},\ }\href {\doibase 10.1080/00031305.1995.10476146} {\bibfield
  {journal} {\bibinfo  {journal} {Amer. Stat.}\ }\textbf {\bibinfo {volume}
  {49}},\ \bibinfo {pages} {217} (\bibinfo {year} {1995})}\BibitemShut
  {NoStop}%
\bibitem [{\citenamefont {Hess}\ \emph {et~al.}(1991)\citenamefont {Hess},
  \citenamefont {Frenkel},\ and\ \citenamefont {Allen}}]{Hess1991}%
  \BibitemOpen
  \bibfield  {author} {\bibinfo {author} {\bibfnamefont {S.}~\bibnamefont
  {Hess}}, \bibinfo {author} {\bibfnamefont {D.}~\bibnamefont {Frenkel}}, \
  and\ \bibinfo {author} {\bibfnamefont {M.~P.}\ \bibnamefont {Allen}},\ }\href
  {\doibase 10.1080/00268979100102561} {\bibfield  {journal} {\bibinfo
  {journal} {Mol. Phys.}\ }\textbf {\bibinfo {volume} {74}},\ \bibinfo {pages}
  {765} (\bibinfo {year} {1991})}\BibitemShut {NoStop}%
\bibitem [{\citenamefont {Dembo}\ and\ \citenamefont
  {Zeitouni}(1986)}]{dembo1986}%
  \BibitemOpen
  \bibfield  {author} {\bibinfo {author} {\bibfnamefont {A.}~\bibnamefont
  {Dembo}}\ and\ \bibinfo {author} {\bibfnamefont {O.}~\bibnamefont
  {Zeitouni}},\ }\href {\doibase 10.1016/0304-4149(86)90018-9} {\bibfield
  {journal} {\bibinfo  {journal} {Stochastic Processes and their Applications}\
  }\textbf {\bibinfo {volume} {23}},\ \bibinfo {pages} {91} (\bibinfo {year}
  {1986})}\BibitemShut {NoStop}%
\bibitem [{\citenamefont {Campillo}\ and\ \citenamefont
  {Gland}(1989)}]{campillo1989}%
  \BibitemOpen
  \bibfield  {author} {\bibinfo {author} {\bibfnamefont {F.}~\bibnamefont
  {Campillo}}\ and\ \bibinfo {author} {\bibfnamefont {F.~L.}\ \bibnamefont
  {Gland}},\ }\href {\doibase 10.1016/0304-4149(89)90041-0} {\bibfield
  {journal} {\bibinfo  {journal} {Stochastic Processes and their Applications}\
  }\textbf {\bibinfo {volume} {33}},\ \bibinfo {pages} {245} (\bibinfo {year}
  {1989})}\BibitemShut {NoStop}%
\bibitem [{\citenamefont {Cauchy}(1829)}]{Cauchy1829}%
  \BibitemOpen
  \bibfield  {author} {\bibinfo {author} {\bibfnamefont {A.~L.}\ \bibnamefont
  {Cauchy}},\ }in\ \href {http://gallica.bnf.fr/ark:/12148/bpt6k90201q/f177}
  {\emph {\bibinfo {booktitle} {Oeuvres complètes d'Augustin Cauchy 2}}},\
  Vol.~\bibinfo {volume} {9}\ (\bibinfo  {publisher} {Gauthier-Villars et
  fils},\ \bibinfo {address} {Paris},\ \bibinfo {year} {1829})\ pp.\ \bibinfo
  {pages} {174--195}\BibitemShut {NoStop}%
\bibitem [{\citenamefont {Fan}\ and\ \citenamefont {Pall}(1957)}]{Fan1957}%
  \BibitemOpen
  \bibfield  {author} {\bibinfo {author} {\bibfnamefont {K.}~\bibnamefont
  {Fan}}\ and\ \bibinfo {author} {\bibfnamefont {G.}~\bibnamefont {Pall}},\
  }\href {\doibase 10.4153/CJM-1957-036-1} {\bibfield  {journal} {\bibinfo
  {journal} {Canad. J. Math.}\ }\textbf {\bibinfo {volume} {9}},\ \bibinfo
  {pages} {298} (\bibinfo {year} {1957})}\BibitemShut {NoStop}%
\bibitem [{\citenamefont {Kingman}(1961)}]{kingman1961}%
  \BibitemOpen
  \bibfield  {author} {\bibinfo {author} {\bibfnamefont {J.~F.~C.}\
  \bibnamefont {Kingman}},\ }\href {\doibase 10.1093/qmath/12.1.283} {\bibfield
   {journal} {\bibinfo  {journal} {Quart. J. Math.}\ }\textbf {\bibinfo
  {volume} {12}},\ \bibinfo {pages} {283} (\bibinfo {year} {1961})}\BibitemShut
  {NoStop}%
\bibitem [{\citenamefont {An}(1998)}]{An1998}%
  \BibitemOpen
  \bibfield  {author} {\bibinfo {author} {\bibfnamefont {M.~Y.}\ \bibnamefont
  {An}},\ }\href {\doibase 10.1006/jeth.1998.2400} {\bibfield  {journal}
  {\bibinfo  {journal} {J. Econ. Theory}\ }\textbf {\bibinfo {volume} {80}},\
  \bibinfo {pages} {350} (\bibinfo {year} {1998})}\BibitemShut {NoStop}%
\bibitem [{\citenamefont {Bagnoli}\ and\ \citenamefont
  {Bergstrom}(2005)}]{Bagnoli2005}%
  \BibitemOpen
  \bibfield  {author} {\bibinfo {author} {\bibfnamefont {M.}~\bibnamefont
  {Bagnoli}}\ and\ \bibinfo {author} {\bibfnamefont {T.}~\bibnamefont
  {Bergstrom}},\ }\href {\doibase 10.1007/s00199-004-0514-4} {\bibfield
  {journal} {\bibinfo  {journal} {Econ. Theor.}\ }\textbf {\bibinfo {volume}
  {26}},\ \bibinfo {pages} {445} (\bibinfo {year} {2005})}\BibitemShut
  {NoStop}%
\bibitem [{\citenamefont {Gupta}\ and\ \citenamefont
  {Balakrishnan}(2012)}]{Gupta2012}%
  \BibitemOpen
  \bibfield  {author} {\bibinfo {author} {\bibfnamefont {R.~C.}\ \bibnamefont
  {Gupta}}\ and\ \bibinfo {author} {\bibfnamefont {N.}~\bibnamefont
  {Balakrishnan}},\ }\href {\doibase 10.1007/s00184-010-0321-9} {\bibfield
  {journal} {\bibinfo  {journal} {Metrika}\ }\textbf {\bibinfo {volume} {75}},\
  \bibinfo {pages} {181} (\bibinfo {year} {2012})}\BibitemShut {NoStop}%
\end{thebibliography}%

\end{document}